\documentclass[12pt,a4wide]{article}
\usepackage{graphicx}% Include figure files
\usepackage{amsmath,amssymb,amsfonts}

\newcommand{\eqa}{\begin{eqnarray}}
\newcommand{\ena}{\end{eqnarray}}
\newcommand{\topstar}[1]{\setlength{\unitlength}{1mm}
\begin{picture}(2,0)(-1,-1.4)
   \put(0,0){\makebox(0,0){$#1$}}
   \put(0,2.4){\makebox(0,0){\mbox{\tiny$\star$}}}
\end{picture}}
\begin{document}
\begin{center}
{\large {\bf Equilibrium Thermodynamics and Neutrino Decoupling
in Quasi-Metric Cosmology}}
\end{center}
\begin{center}
Dag {\O}stvang \\
{\em Department of Physics, Norwegian University of Science and Technology
(NTNU), \\
N-7491 Trondheim, Norway}
\end{center}
\begin{abstract}
The laws of thermodynamics in the expanding universe are formulated within the
quasi-metric framework. Since the quasi-metric cosmic expansion does not 
directly influence momenta of material particles, so that the expansion 
directly cools the photons only (or more generally, null particles), these laws
differ substantially from their counterparts in standard cosmology. An 
approximate model describing thermodynamics during neutrino decoupling is set 
up. This model yields, assuming that no neutrino mass eigenstate is null so 
that the expansion does not directly cool the neutrinos, that the result after 
neutrino decoupling will be a non-thermal relic neutrino background which 
decoupled gradually over about 20-30 years when the photon plasma had a 
temperature of ${\sim}{\rm 100-50{\ }keV}/k_{\rm B}$. The relic neutrinos are 
predicted to have a number density today more than twice the standard cosmology
result and to have the same energy distribution today as they had just after 
decoupling. As a consequence of this, the relic neutrino background is 
predicted to consist of neutrino mass eigenstates with an average energy of 
${\sim}{\rm 170{\ }keV}$. This predicted relic neutrino background is strongly 
inconsistent with detection rates measured in solar neutrino detectors 
(Borexino in particular), unless some particular property of neutrinos 
makes the relic neutrino background essentially undetectable
(e.g., if one neutrino mass eigenstate is null and the two massive mass 
eigenstates decay into the the null eigenstate over cosmic time scales).
But in the absence of such a natural explanation from neutrino physics, the 
current status of quasi-metric relativity has been changed to non-viable.
\\
\end{abstract}
\topmargin 0pt
\oddsidemargin 5mm
\renewcommand{\thefootnote}{\fnsymbol{footnote}}
\section{Introduction}
Nowadays one often hears the assertion that the science of cosmology has 
matured to the point where one speaks of ``precision cosmology'', meaning that
its theoretical foundations are considered beyond reasonable doubt and that 
almost all cosmological data are consistent with the standard big bang (SBB) 
scenario (currently characterized by a positive cosmological constant and cold 
dark matter) based on general relativity (GR). Reasons for this view are based 
on analyses of the cosmic microwave background (CMB) and moreover on the 
predictions coming from standard big bang nucleosynthesis (SBBN) agreeing well 
with the observed abundances of light elements, so that together with results 
coming from other cosmological observations, the values of the cosmological 
parameters can be inferred in a consistent way. In particular the successes of
SBBN are convincing since its predictions are overconstrained and not a result
of mere parameter-fitting exercises.

However, mainstream cosmology is based on GR, so the successes of the SBB
scenario cannot be independent of how well GR fares on smaller scales. 
Unfortunately, GR has difficulties explaining galactic phenomenology such as
rotation curve shapes without doing more than essentially fitting suitable
dark matter distributions to give the desired results. Moreover, the physics
behind observed scaling relations such as the Tully-Fisher relation has no
obvious theoretical basis in GR (or in Newtonian gravity) whatsoever. Even on
such a small scale as the solar system there are anomalies not well explained
by GR (or Newtonian gravity). But some of these anomalies have simple 
explanations based on first principles coming from an alternative space-time 
framework, the so-called quasi-metric framework (QMF) [1, 2]. Said explanations
are mostly based on the most characteristic feature of the QMF, namely that 
the cosmic expansion is postulated to be independent of space-time's causal 
structure. As a consequence, the QMF predicts that the cosmic expansion should 
be detectable in the solar system, and that this naturally explains several 
anomalies [3]. This means that there is observational support for the 
possibility that GR mismodels the cosmic expansion, so that some of the 
fundamental assumptions underlying the SBB scenario could be false. This 
indicates that the SBB scenario may not be very robust after all and that any 
talk of ``precision cosmology'' would be premature.

But even if the QMF is able to explain some solar system anomalies in addition
to the usual gravitational solar system tests, it has not yet been shown to be 
viable in general. One step towards viability would be to show that the
predicted thermodynamics of the early Universe is consistent with current
observational results. This goal motivates further exploration of quasi-metric 
cosmology and the question of its viability, which is the theme of the present 
paper. However, as we shall see, the results are not promising since the 
properties of the predicted cosmic neutrino background are in violent conflict 
with observations.

In short, the crucial point is that in the QMF, the momenta of material 
particles are not directly affected by the cosmic expansion. This is a unique 
feature of the QMF and as a consequence, thermodynamics in the expanding 
universe as described within the QMF, is nonstandard. This results in the 
specific prediction that there should at present exist a non-thermal relic 
cosmological neutrino background in the form of neutrino mass eigenstates, with
average energy of $\sim$ 170 keV. Unless some exotic properties of neutrinos 
make the relic neutrino background essentially unobservable (e.g., if one 
neutrino mass eigenstate is null and the two massive mass eigenstates decay 
into the the null eigenstate over cosmic time scales), said prediction is in 
violent conflict with measured results from solar neutrino observatories such 
as SAGE, GALLEX/GNO and Borexino. In view of this result the status of the QMF 
has been changed to not viable.
\section{Motivating quasi-metric relativity}
The quasi-metric space-time framework (QMF), along with motivations for 
introducing it, has been published in [1]. A synopsis is presented in the
present paper for the benefit of new readers.

The main motivation for inventing the QMF is of a very general philosophical 
nature. That is, traditional field theories consist of two independent parts; 
field equations and initial conditions. This form ensures that field theories 
can in principle be generally applied to all systems within their domain of 
validity. But for cosmology this flexibility is a liability; since the Universe
is unique, it is impossible in principle to have observational knowledge of 
alternatives to cosmic initial conditions, global evolution and structure. Any 
diversity of such possibilities represents a serious limitation to what can be 
known in principle, and should be avoided if possible. That is, since the 
Universe is observationally unique, so should the nature of its global 
evolution be. 

It turns out that, to construct a general framework fulfilling this 
requirement, one is pretty much led to the QMF. Moreover, the QMF accomplishes 
this requirement by describing the global cosmic expansion as an absolute, 
prior-geometric phenomenon, not being part of space-time's causal structure. 
In this way the cosmic expansion does not depend on field equations and
initial conditions, meaning that the Universe is not described as a purely
dynamical system.

Similar to the Robertson-Walker (RW) models in GR, the cosmic expansion in the 
QMF is defined by means of a family of ``preferred'' observers, the so-called
{\em fundamental observers} (FOs). A further similarity with the RW-models is 
the existence of a {\em global time function} $t$, such that $t$ splits up 
space-time into a ``distinguished'' set of  spatial hypersurfaces, the so-called
{\em fundamental hypersurfaces} (FHSs). But since the cosmic expansion in the 
QMF by hypothesis is not part of space-time's causal structure, $t$ cannot be 
an ordinary time coordinate on a Lorentzian manifold. Rather, it should play 
the role of an independent evolution parameter, parametrizing any change in the 
space-time geometry that has to do with the cosmic expansion. On the other 
hand, space-time must also be equipped with a causal structure in the form of a
Lorentzian manifold. This Lorentzian manifold must also accommodate the FOs and
the FHSs, which means that its topology should allow the existence of a global 
ordinary time coordinate $x^0$. (Note that, to ensure the uniqueness of this 
construction, the FHSs must be compact.)

Taking into account the above considerations, the geometrical basis of the QMF 
can now be defined. That is, the geometry underlying the QMF consists of a 
5-dimensional differentiable manifold with topology ${\cal M}{\times}
{\bf R}_1$, where ${\cal M}={\cal S}{\times}{\bf R}_2$ is a Lorentzian
space-time manifold, ${\bf R}_1$ and ${\bf R}_2$ both denote the real line
and ${\cal S}$ is a compact 3-dimensional manifold (without boundaries). That 
is, in addition to the usual time dimension and 3 space dimensions there is an
extra degenerate time dimension ${\bf R}_1$ represented by the global time 
function $t$. Moreover, the manifold ${\cal M}{\times}{\bf R}_1$ is equipped 
with two 5-dimensional degenerate metrics ${\bf {\bar g}}_t$ and ${\bf g}_t$, 
where the degeneracies are described by the conditions ${\bf {\bar g}}_t
({\frac{\partial}{{\partial}t}},{\cdot}){\equiv}0$ and ${\bf g}_t
({\frac{\partial}{{\partial}t}},{\cdot}){\equiv}0$, respectively. The metric
${\bf {\bar g}}_t$ is coupled to matter fields via field equations, and from
this one can construct the ``physical'' metric ${\bf g}_t$ used when comparing
predictions to experiments. Note that these metrics have the property that
the FOs always move orthogonally to the FHSs.

To reduce space-time to 4 dimensions, one obtains the quasi-metric space-time 
manifold $\cal N$ by slicing the submanifold determined by the equation 
$x^0=ct$ out of the 5-dimensional differentiable manifold ${\cal M}{\times}
{\bf R}_1$. It is essential that this slicing is unique since the two global
time coordinates should be physically equivalent; the only reason to separate
between them is that they are designed to parameterize fundamentally different
physical phenomena. Since the geometric structure on ${\cal N}$ is inherited
from that on ${\cal M}{\times}{\bf R}_1$ just by restricting the fields to
${\cal N}$ (no projections), the 5-dimensional degenerate metric fields 
${\bf {\bar g}}_t$ and ${\bf g}_t$ may be regarded as one-parameter families of
Lorentzian 4-metrics on $\cal N$ (this terminology is merely a matter of 
semantics). Note that there exists a set of particular coordinate systems 
especially well adapted to the geometrical structure of quasi-metric 
space-time, {\em the global time coordinate systems} (GTCSs). A coordinate 
system is a GTCS iff the time coordinate $x^0$ is related to $t$ via $x^0=ct$ 
in ${\cal N}$. 

The role of $t$ in ${\bf {\bar g}}_t$ and ${\bf g}_t$ is to describe how the 
cosmic expansion influences space-time geometry. The {\em global} part of this 
$t$-dependence should enter ${\bf {\bar g}}_t$ and ${\bf g}_t$ explicitly as a 
scale factor. However, unlike its counterpart in the RW-models, this scale 
factor cannot be calculated from dynamical equations, but must be an 
``absolute'' quantity. Since the form of the scale factor should not introduce 
any extra arbitary scale or parameter, the only possibile option for a scale 
factor with the dimension of length is to set it equal to $ct$. In 
$({\cal N},{\bf {\bar g}}_t)$, this scale factor may be multiplied by a second,
dimensionless scale factor taking into account basic effects of gravity. But 
since the basic geometry of the FHSs in $({\cal N},{\bf {\bar g}}_t)$ is 
postulated to represent a measure of gravitational scales in terms of atomic 
units, any extra dimensionless scale factor should enter ${\bf {\bar g}}_t$ as 
a conformal factor. Furthermore, since there is no reason to introduce any 
nontrivial spatial topology, the global basic geometry of the FHSs (neglecting 
the effects of gravity) should be that of the 3-sphere ${\bf S^3}$. However, 
any restriction on the global geometry of the FHSs implies the existence of 
prior 3-geometry. This prior 3-geometry should not restrict the general form of
the metric family explicitly though (but see equation (13) below).
 
Now the general form of the metric family $({\cal N},{\bf {\bar g}}_t)$ can be
represented by the family of line elements ${\overline {ds}}_t^2$, where the 
global dependence on $t$ is included explicitly via the scale factor mentioned 
above. That is, expressed in a suitable GTCS, the most general form allowed for
the family ${\bf {\bar g}}_t$ may be represented by the family of line elements
(we use the metric signature $(-+++)$ and Einstein's summation convention 
throughout)
\eqa
{\overline {ds}}_t^2={\bar N}_t^2{\Big \{ }
[{\bar N}_{(t)}^k{\bar N}_{(t)}^s{\tilde h}_{(t)ks}-1](dx^0)^2+
2{\frac{t}{t_0}}{\bar N}_{(t)}^k{\tilde h}_{(t)ks}dx^sdx^0+
{\frac{t^2}{t_0^2}}{\tilde h}_{(t)ks}dx^kdx^s{\Big \} }.
\ena
(Latin coordinate indices take values in the range 1..3.)
Here, $t_0$ is some arbitrary reference epoch (usually chosen to be the present
epoch) setting the scale of the spatial coordinates, ${\bar N}_t$ is the family
of lapse functions (of the FOs) and ${\frac{t_0}{t}}{\bar N}^k_{(t)}$ are the 
components of the shift vector family (of the FOs) in 
$({\cal N},{\bf {\bar g}}_t)$. Also, ${\bar h}_{(t)ks}{\equiv}{\frac{t^2}{t_0^2}}
{\bar N}_t^2{\tilde h}_{(t)ks}$ are the components of the spatial metric family 
${\bf {\bar h}}_t$ intrinsic to the FHSs. We notice that ${\bar N}_t^2$ enters 
equation (1) as a conformal factor, but that this does not imply any 
restrictions on the general form of ${\bf {\bar g}}_t$. The only point of 
introducing the conformal factor is that the total scale factor of the FHSs may
now be defined formally as ${\bar F}_t{\equiv}{\bar N}_tct$. We also notice 
that ${\bar N}_t$ may depend on $t$ and that the form (1) of ${\bf {\bar g}}_t$
is preserved only under coordinate transformations between GTCSs. 

Next, $({\cal N},{\bf {\bar g}}_t)$ and $({\cal N},{\bf g}_t)$ are equipped 
with linear and symmetric connections ${\ }{\topstar{\bf {\bar {\nabla}}}}{\ }$
and ${\ }{\topstar{\bf {\nabla}}}{\ }$, respectively. These connections are
identified with the usual Levi-Civita connection for constant $t$, yielding the
standard form of the connection coefficients not containing $t$. The rest of 
the connection coefficients are determined by the requirements
\eqa
{\topstar{\bf {\bar {\nabla}}}}_{\frac{\partial}{{\partial}t}}
{\bf {\bar g}}_t=0, \qquad
{\topstar{\bf {\bar {\nabla}}}}_{\frac{\partial}{{\partial}t}}
{\bf {\bar n}}_t=0, \qquad
{\topstar{\bf {\nabla}}}_{\frac{\partial}{{\partial}t}}
{\bf g}_t=0, \qquad
{\topstar{\bf {\nabla}}}_{\frac{\partial}{{\partial}t}}
{\bf n}_t=0,
\ena
where ${\bf {\bar n}}_t$ and ${\bf n}_t$ are families of unit normal vector 
fields to the FHSs in $({\cal N},{\bf {\bar g}}_t)$ and $({\cal N},{\bf g}_t)$,
respectively. Note that we have the general decomposition formula
\eqa
{\bf g}_t=-{\bf g}_t({\bf n}_t,{\cdot}){\otimes}
{\bf g}_t({\bf n}_t,{\cdot})+{\bf h}_t,
\ena
(and a counterpart formula valid for ${\bf {\bar g}}_t$), where ${\bf h}_t$ is 
the metric family intrinsic to the FHSs in $({\cal N},{\bf g}_t)$. This implies
that we also must have that ${\,}
{\topstar{\bf {\bar {\nabla}}}}_{\frac{\partial}{{\partial}t}}{\bf {\bar h}}_t=0$ 
and ${\,}{\topstar{\bf {\nabla}}}_{\frac{\partial}{{\partial}t}}{\bf h}_t=0$. 

The requirements shown in equation (2) yield the nonzero extra connection 
coefficients (using a GTCS and where a comma denotes a partial derivative)
\eqa
{\topstar{\bar {\Gamma}}}^{{\,}0}_{t0}={\frac{{\bar N}_t,_t}{{\bar N}_t}}, 
\qquad {\topstar{\bar {\Gamma}}}_{tj}^{{\,}i}=
{\Big (}{\frac{1}{t}}+{\frac{{\bar N}_{t,t}}{{\bar N}_t}}{\Big )}{\delta}^i_j
+{\frac{1}{2}}{\tilde h}_{(t)}^{is}{\tilde h}_{(t)sj,t}, 
\qquad {\topstar {\Gamma}}^{{\,}i}_{jt}={\frac{1}{2}}h_{(t)}^{is}
{\frac{\partial}{{\partial}t}}h_{(t)sj},
\ena
in addition to identical expressions for those obtained by permuting the two 
lower indices.

As mentioned earlier, the scale factor ${\bar N}_tct$ of the FHSs as obtained 
from equation (1), is interpreted as a gravitational scale measured in atomic 
units. This interpretation must also hold for all dimensionful gravitational 
quantities with dimension of length to some power (here, time scales the same 
way as length and inversely of mass while charge is in effect dimensionless). 
In particular this applies to the gravitational coupling parameter $G_t$, 
effectively scaling as length squared. However, $G_t$ couples both to mass and 
to charge squared, so if one wants to transfer the variability of $G_t$ to
matter sources, it will be necessary to define {\em two} gravitational 
constants $G^{\rm B}$ and $G^{\rm S}$. Here, $G^{\rm B}$ is the ``bare'' 
gravitational constant coupling to charge or more generally to the 
electromagnetic field and in principle measurable in local gravitational 
experiments at an arbitrary reference event epoch $t_0$. Moreover, $G^{\rm S}$ 
is the ``screened'' gravitational constant coupling to material matter fields 
and in principle also measurable in local gravitational experiments at the 
reference epoch $t_0$. This means that we must have non-universal 
gravitational coupling between matter sources and space-time geometry. The
necessity of having non-universal gravitational coupling was missed in the
original formulation of the QMF.

The variability of dimensionful gravitational quantities as measured in atomic
units, in addition to transferring the variability of $G_t$ to matter sources,
imply that one must distinguish between {\em active mass} measured dynamically 
as a source of gravity and {\em passive mass}, i.e., passive gravitational mass
or inertial mass. (Similarly one must distinguish between {\em active charge} 
and {\em passive charge}.) Taking into account said variation of gravitational 
scales measured in atomic units, it is possible to set up local conservation 
laws in $({\cal N},{\bf {\bar g}}_t)$ involving the covariant derivative 
(holding $t$ fixed) ${\bf {\bar {\nabla}}}{\bf {\cdot}}{\bf T}_t$ of the 
{\em active stress-energy tensor} ${\bf T}_t$. These local conservation laws 
do not depend on the nature of the source. In component notation, they take 
the form [1, 2] (greek coordinate indices take values in the range 0..3)
\eqa
T^{\nu}_{(t){\mu};{\nu}}=
2{\frac{{\bar N}_{t,{\nu}}}{{\bar N}_t}}T^{\nu}_{(t){\mu}}
=2c^{-2}{\bar a}_{{\cal F}s}{\hat T}^s_{(t){\mu}}
-2{\frac{{\bar N}_t,_{\bar {\perp}}}{{\bar N}_t}}T_{(t){\bar {\perp}}{\mu}},
\qquad c^{-2}{\bar a}_{{\cal F}s}{\equiv}{\frac{{\bar N}_{t,s}}{{\bar N}_t}}.
\ena
Here, the projection symbol '${\bar {\perp}}$' denotes a scalar product with 
$-{\bf {\bar n}}_t$ and the ``hat'' symbol marking above a space-time object 
denotes that upper space indices are raised with the inverse space metric 
family ${\bf {\bar h}}_t^{-1}$ rather than with the inverse space-time metric 
family ${\bf {\bar g}}_t^{-1}$. The local conservation laws (5) imply that 
inertial test particles move along geodesics of 
${\ }{\topstar{\bf {\bar {\nabla}}}}{\ }$ in $({\cal N},{\bf {\bar g}}_t)$, 
and this ensures that inertial test particles move along geodesics of
${\ }{\topstar{\bf {\nabla}}}{\ }$ in $({\cal N},{\bf g}_t)$ as well [1, 2].
Note that due to equation (2), the lapse function $N$ of the FOs in 
$({\cal N},{\bf g}_t)$ cannot depend explicitly on $t$. This means that, when 
making the transformation ${\bf {\bar g}}_t{\rightarrow}{\bf g}_t$, $N$ gets an
``effective'' time dependence parametrized by $x^0$. Then the equations of 
motion in $({\cal N},{\bf g}_t)$ read (in a GTCS, in component notation)
\eqa
{\frac{d^2x^{\mu}}{d{\lambda}^2}}+{\Big (}
{\topstar{\Gamma}}_{t{\nu}}^{{\,}{\mu}}{\frac{dt}{d{\lambda}}}+
{\topstar{\Gamma}}_{{\beta}{\nu}}^{{\,}{\mu}}{\frac{dx^{\beta}}{d{\lambda}}}
{\Big )}{\frac{dx^{\nu}}{d{\lambda}}}
={\Big (}{\frac{d{\tau}_t}{d{\lambda}}}{\Big )}^2a_{(t)}^{\mu},
\ena
where $d{\tau}_t$ is the proper time interval as measured along the curve,
${\lambda}$ is some general affine parameter, and ${\bf a}_t$ is the 
4-acceleration measured along the curve.

Due to the need for non-universal gravitational couplings, it is convenient to
split up ${\bf T}_t$ into one electromagnetic part ${\bf T}_t^{\rm (EM)}$ and one
part ${\bf T}_t^{\rm (MA)}$ representing matter fields. However, it turns out [1] 
that full couplings of ${\bf T}^{\rm (EM)}_t$ and ${\bf T}^{\rm (MA)}_t$ to 
space-time curvature cannot exist. Moreover, only that part of the space-time 
curvature obtained by setting $t$ constant should couple to matter. 
Fortunately, one of the projected (generalized) Einstein field equations can be
tailored to ${\bf {\bar g}}_t$ so that {\em partial} couplings exist, yielding 
approximate correspondences with GR and Newtonian gravity for isolated systems 
and weak fields. Said projected field equation reads (expressed in a GTCS)
\eqa
{\bar R}_{(t){\bar {\perp}}{\bar {\perp}}}=c^{-4}{\bar a}_{{\cal F}k}
{\bar a}_{\cal F}^k+c^{-2}{\bar a}^k_{{\cal F}{\mid}k}-{\bar K}_{(t)ik}
{\bar K}_{(t)}^{ik}+{\pounds}_{{\bf {\bar n}}_t}{\bar K}_t \nonumber \\
={\frac{4{\pi}G^{\rm B}}{c^4}}(T^{\rm (EM)}_{(t){\bar {\perp}}{\bar {\perp}}}
+{\hat T}^{{\rm (EM)}i}_{(t)i})+{\frac{4{\pi}G^{\rm S}}{c^4}}
(T_{(t){\bar {\perp}}{\bar {\perp}}}^{\rm (MA)}+{\hat T}^{{\rm (MA)}i}_{(t)i}),
\ena
where ${\bf {\bar R}}_t$ is the Ricci tensor family and ${\bf {\bar K}}_t$
is the extrinsic curvature tensor family (with trace ${\bar K}_t$) of the FHSs 
obtained from equation (1). Moreover, the symbol '${\mid}$' denotes a spatial
covariant derivative obtained from the connection intrinsic to the FHSs and 
the operation ${\pounds}_{{\bf {\bar n}}_t}$ denotes a Lie derivative in the 
${{\bf {\bar n}}_t}$-direction. 

A second set of field equations having some similarity (apart from the extra 
terms) to other relevant projections of the (generalized) Einstein field 
equations is (in a GTCS)
\eqa
{\bar R}_{(t)j{\bar {\perp}}}+{\Big (}{\frac{{\bar h}_{(t)}^{ik}}{{\bar N}_t}}
{\frac{\partial}{{\partial}x^0}}{\bar h}_{(t)ij}{\Big )}_{{\mid}k}-
{\Big (}{\frac{{\bar h}_{(t)}^{ik}}{{\bar N}_t}}
{\frac{\partial}{{\partial}x^0}}{\bar h}_{(t)ik}{\Big )},_j
={\frac{8{\pi}G^{\rm B}}{c^4}}T^{{\rm (EM)}}_{(t)j{\bar {\perp}}}
+{\frac{8{\pi}G^{\rm S}}{c^4}}T^{\rm (MA)}_{(t)j{\bar {\perp}}}.
\ena
Finally, a third set of field equations having no similarities at all to the 
counterpart projections of the Einstein field equations reads 
\eqa
{\bar C}_{(t){\bar {\perp}}i{\bar {\perp}}j}={\tilde H}_{(t)ij}+
{\frac{1}{(ct{\bar N}_t)^2}}{\bar h}_{(t)ij},
\ena
where ${{\bf {\bar C}}_t}$ is the Weyl tensor family in 
$({\cal N},{\bf {\bar g}}_t)$ and ${\bf {\tilde H}}_t$ is the spatial Einstein 
tensor family calculated from the metric family ${\bf {\tilde h}}_t{\equiv}
{\frac{t_0^2}{t^2}}{\bar N}_t^{-2}{\bf {\bar h}}_t$. Note that in equation (9), 
there is no explicit coupling present to the relevant projections of 
${\bf T}_t$. Also, note in particular that equations (7)-(9) are valid only for
the foliation of $({\cal N},{\bf {\bar g}}_t)$ into the set of FHSs. For 
foliations of $({\cal N},{\bf {\bar g}}_t)$ into other (spatial) hypersurfaces 
they will not be valid in general. (This can be easily seen from equations (8) 
and (9); it is not possible that these could be valid for arbitrary foliations 
of $({\cal N},{\bf {\bar g}}_t)$ into spatial hypersurfaces since these 
equations involve foliation-dependent quantities that cannot be represented by 
projections of space-time tensor families.)

An explicit coordinate expression for ${\bf {\bar K}}_t$ is given from the 
well-known (except for the $t$-dependence) formula from canonical GR
\eqa
{\bar K}_{(t)ij}={\frac{1}{2{\bar N}_t}}{\Big [}{\frac{t}{t_0}}
({\bar N}_{(t)i{\mid}j}+{\bar N}_{(t)j{\mid}i})-
{\frac{\partial}{{\partial}x^0}}{\bar h}_{(t)ij}{\Big ]},
\ena
\eqa
{\bar K}_t={\frac{t_0}{t}}{\frac{{\bar N}^i_{(t){\mid}i}}{{\bar N}_t}}
-{\frac{1}{2{\bar N}_t}}{\bar h}_{(t)}^{ij}{\frac{\partial}{{\partial}x^0}}
{\bar h}_{(t)ij}.
\ena
It is also convenient to have explicit expressions for the curvature intrinsic 
to the FHSs. From equation (1) one easily calculates (given the 
prior-geometric restriction on ${\tilde P}_t$, see equation (13) below)
\eqa
{\bar H}_{(t)ij}=c^{-2}{\Big (}{\bar a}_{{\cal F}{\mid}k}^k- 
{\frac{1}{{\bar N}_t^2t^2}}{\Big )}{\bar h}_{(t)ij}-c^{-4}
{\bar a}_{{\cal F}i}{\bar a}_{{\cal F}j}-c^{-2}
{\bar a}_{{\cal F}i{\mid}j}+{\tilde H}_{(t)ij},
\ena
\eqa
{\bar P}_t={\frac{6}{({\bar N}_tct)^2}}+2c^{-4}{\bar a}_{{\cal F}k}
{\bar a}_{\cal F}^k-4c^{-2}{\bar a}_{{\cal F}{\mid}k}^k, \qquad
{\tilde P}_t={\frac{6}{(ct_0)^2}},
\ena
where ${\bf {\bar H}}_t$ is the Einstein tensor family intrinsic to the FHSs
in $({\cal N},{\bf {\bar g}}_t)$ and ${\bar P}_t$ is the corresponding 
curvature scalar family. Similarly, ${\tilde P}_t$ is the curvature scalar
associated with the metric family ${\bf {\tilde h}}_t$. Note that the form of
${\tilde P}_t$ represents the prior 3-geometry of the FHSs and that it does not 
depend explicitly on $t$ once an arbitrary reference epoch $t_0$ has been 
chosen.
\section{The early quasi-metric universe}
\subsection{Cosmological space-time geometry}
As indicated by observations, the early Universe was highly isotropic and
homogeneous. This means that to model the early quasi-metric universe,
using a spherical GTCS ${\{}x^0,{\chi},{\theta},{\phi}{\}}$, equation (1) 
should take the form
\eqa
{\overline {ds}}^2_t={\bar N}_t^2{\Big \{}-(dx^0)^2+
(ct)^2{\Big (}d{\chi}^2+{\sin}^2{\chi}d{\Omega}^2{\Big )}{\Big {\}}},
\ena
where $d{\Omega}^2{\equiv}d{\theta}^2+{\sin}^2{\theta}d{\phi}^2$ is the solid
angle line element and where ${\bar N}_t$ does not depend on the spatial 
coordinates. The simplest case of the line element family (14) is when 
${\bar N}_t$ is a constant in ${\cal M}{\times}{\bf R}_1$; this is identified 
as a vacuum solution since equations (5), (7), (8) and (9) are then trivially 
fulfilled. In fact, a vacuum is the natural beginning of the quasi-metric 
universe since this avoids any problems with diverging physical quantities at 
the initial geometrical singularity. But this means that one needs to have a 
matter creation mechanism at work in the very early quasi-metric universe,
presumably expressed via the $t$-dependence of ${\bf T}_t$ so that equation
(5) should still be valid. However, this part of quasi-metric cosmology has not
yet been developed, and is beyond the scope of the present paper. In the rest 
of this paper it is assumed that effects of net matter creation can be 
neglected.

We now consider the quasi-metric universe filled with matter modelled as a
perfect fluid (comoving with the FOs) with active mass density 
${\tilde {\varrho}}_{\rm m}$ and active pressure ${\tilde p}$, i.e.,
\eqa
T_{(t){\bar {\perp}}{\bar {\perp}}}={\tilde {\varrho}}_{\rm m}c^2{\equiv}
{\Big (}{\frac{t_0}{{\bar N}_tt}}{\Big )}^2{\bar {\varrho}}_{\rm m}c^2,
\qquad T_{(t){\chi}}^{\chi}=T_{(t){\theta}}^{\theta}=T_{(t){\phi}}^{\phi}=
{\tilde p}{\equiv}
{\Big (}{\frac{t_0}{{\bar N}_tt}}{\Big )}^2{\bar p},
\ena
where we have set the (arbitrary) boundary condition ${\bar N}_t(t_0)=1$ in 
$({\cal N},{\bf {\bar g}}_t)$ for the arbitrary reference epoch $t_0$.
Moreover, ${\bar {\varrho}}_{\rm m}$ is by definition the {\em properly scaled 
density of active mass} and ${\bar p}$ is the associated pressure. The 
condition that there is no net matter creation is expressed as 
${\frac{\partial}{{\partial}t}}{\bar {\varrho}}_{\rm m}=0$. Also, we have the 
relationship
\eqa
{\bar {\varrho}}_{\rm m}=
\left\{
\begin{array}{ll}
{\frac{t^3}{t_0^3}}{\bar N}_t^3{\varrho}_{\rm m}
& \text{for a fluid of material particles,} \\ [1.5 ex]
{\frac{t^4}{t_0^4}}{\bar N}_t^4{\varrho}_{\rm m} & \text{for the 
electromagnetic field,}
\end{array}
\right.
\ena
between ${\bar {\varrho}}_{\rm m}$ and the directly measurable passive 
(inertial) mass density ${\varrho}_{\rm m}$. An identical relationship exists 
between ${\bar p}$ and the passive pressure $p$. Besides, projecting equation 
(5) with respect to the FHSs (see, e.g., [1] for explicit formulae) and using 
equation (15) then yields
\eqa
{\pounds}_{{\bf {\bar n}}_t}T_{(t){\bar {\perp}}{\bar {\perp}}}=
{\frac{t_0^2}{t^2}}{\frac{{\bar N}_{t,{\bar {\perp}}}}{{\bar N}_t}}
{\Big (}{\bar {\varrho}}_{\rm m}c^2+3{\bar p}{\Big )}, \quad \Rightarrow \quad
{\pounds}_{{\bf {\bar n}}_t}{\bar {\varrho}}_{\rm m}=
-{\frac{{\bar N}_{t,{\bar {\perp}}}}{{\bar N}_t}}
{\Big (}{\bar {\varrho}}_{\rm m}-3{\bar p}/c^2{\Big )},
\ena
where the last expression must vanish since no other possibilities exist
satisfying the field equation (7) with a proper vacuum limit. This means
that ${\bar {\varrho}}_{\rm m}$ does not depend on $x^0$, so to model a 
nonvacuum isotropic and homogeneous quasi-metric universe, the fluid must 
satisfy the equation of state ${\varrho}_{\rm m}=3p/c^2$, i.e., it must be a 
null fluid. A material fluid can only be considered if it is so hot that any 
deviation from said equation of state is utterly negligible, so that any 
corresponding deviation from isotropy can also be neglected. A hot plasma
consisting mainly of photons and neutrinos (with negligible amounts of more 
massive particles) in thermal equilibrium, as found at the later epochs of a
radiation-dominated universe, will satisfy this condition to a good 
approximation.

To show that an isotropic and homogeneous universe filled with a null fluid is 
indeed possible in quasi-metric gravity, we must solve the field equation (7) 
using equations (10), (11) and (14) (equations (8) and (9) are trivially 
fulfilled). We then get
\eqa
{\pounds}_{{\bf {\bar n}}_t}{\Big (}
{\frac{{\bar N}_{t,{\bar {\perp}}}}{{\bar N}_t}}{\Big )}-
{\Big (}{\frac{{\bar N}_{t,{\bar {\perp}}}}{{\bar N}_t}}{\Big )}^2 \nonumber \\
={\frac{4{\pi}G^{\rm B}}{3c^4}}{\frac{t_0^2}{{\bar N}_t^2t^2}}
{\Big (}{\bar {\varrho}}^{\rm (EM)}_{\rm m}c^2+3{\bar p}^{\rm (EM)}{\Big )}+
{\frac{4{\pi}G^{\rm S}}{3c^4}}{\frac{t_0^2}{{\bar N}_t^2t^2}}
{\Big (}{\bar {\varrho}}^{\rm (MA)}_{\rm m}c^2+3{\bar p}^{\rm (MA)}{\Big )},
\ena
or equivalently, for a null fluid,
\eqa
{\Big (}{\frac{{\bar N}_{t,0}}{{\bar N}_t}}{\Big )}_{,0}=
-{\frac{8{\pi}G^{\rm B}}{3c^4}}{\frac{t_0^2}{t^2}}{\bar {\varrho}}^{\rm (EM)}
_{\rm m}c^2-{\frac{8{\pi}G^{\rm S}}{3c^4}}{\frac{t_0^2}{t^2}}
{\bar {\varrho}}^{\rm (MA)}_{\rm m}c^2,
\ena
where ${\bar {\varrho}}_{\rm m}={\bar {\varrho}}^{\rm (EM)}_{\rm m}+
{\bar {\varrho}}^{\rm (MA)}_{\rm m}$ has been split up into contributions
${\bar {\varrho}}^{\rm (EM)}_{\rm m}$ and ${\bar {\varrho}}^{\rm (MA)}_{\rm m}$
arising from electromagnetic sources and from material particle sources, 
respectively. Integrating equation (19) twice and requiring consistency with 
the chosen boundary condition, and furthermore requiring continuity with the 
empty solution in the vacuum limit ${\bar {\varrho}}_{\rm m}{\rightarrow}0$, we 
find the solution
\eqa
{\bar N}_t={\exp}{\Big [}-{\frac{4{\pi}G^{\rm B}}{3}}t_0^2{\Big (}
{\frac{(x^0)^2}{(ct)^2}}{\bar {\varrho}}^{\rm (EM)}_{\rm m}(t)
-{\bar {\varrho}}^{\rm (EM)}_{\rm m}(t_0){\Big )} \nonumber \\
-{\frac{4{\pi}G^{\rm S}}{3}}t_0^2{\Big (}{\frac{(x^0)^2}{(ct)^2}}
{\bar {\varrho}}^{\rm (MA)}_{\rm m}(t)
-{\bar {\varrho}}^{\rm (MA)}_{\rm m}(t_0){\Big )}{\Big ]}.
\ena
We see from equation (20) that in $({\cal N},{\bf {\bar g}}_t)$, ${\bar N}_t>1$
for very early epochs, were there is net matter creation 
${\frac{\partial}{{\partial}t}}{\bar {\varrho}}_{\rm m}>0$. Then ${\bar N}_t$
decreases towards unity until the epoch where matter creation becomes 
negligible, so that for later epochs, ${\bar {\varrho}}_{\rm m}(t)=
{\bar {\varrho}}_{\rm m}(t_0)$. Since we have assumed the latter, we may thus 
set ${\bar N}_t=1$ in $({\cal N},{\bf {\bar g}}_t)$ for the rest of this paper.

Actually ${\frac{\partial}{{\partial}t}}{\bar {\varrho}}^{\rm (EM)}_{\rm m}>0$ and
${\frac{\partial}{{\partial}t}}{\bar {\varrho}}^{\rm (MA)}_{\rm m}<0$ due to net
heat transfer between photons and material particles in thermal equilibrium.
This means that ${\bar N}_t{\neq}1$ in $({\cal N},{\bf {\bar g}}_t)$ even in 
absence of matter creation. However, for the late stages of the 
radiation-dominant epoch, only neutrinos are abundant enough to make this 
effect significant, so ${\bar N}_t$ will still have no spatial dependence to a 
very good approximation. This means that setting ${\bar N}_t=1$ in 
$({\cal N},{\bf {\bar g}}_t)$ does not affect significantly the main results
of this paper.
\subsection{Particle kinematics in quasi-metric spacetime}
We will now consider geodesic motion of particles from equation (6) and 
calculate how the cosmic expansion affects a particle's speed and momentum. As 
we shall see, for material particles the results will be crucially different 
from their counterparts in GR.

To begin with, we notice that the transformation ${\bf {\bar g}}_t{\rightarrow}
{\bf g}_t$ (see, e.g., [1]) reduces to setting ${\bar N}_t=N$ for the special 
case where ${\bf {\bar a}}_{\cal F}{\equiv}0$, valid for the familiy 
of line elements given in equation (14). Thus the family of line elements 
corresponding to the metric family $({\cal N},{\bf g}_t)$ can be found from 
equation (14) simply by setting ${\bar N}_t=1$. Next, rewrite equation (6) for 
geodesic motion in terms of the family of 4-velocities ${\bf u}_t$ (with 
components $u_{(t)}^{\mu}={\frac{dx^{\mu}}{d{\tau}_t}}$) of a material 
particle by setting ${\lambda}=c{\tau}_t$. Equation (6) then takes the form
\eqa
{\frac{du_{(t)}^{\mu}}{d{\tau}_t}}+{\Big (}
{\topstar{\Gamma}}_{t{\nu}}^{{\,}{\mu}}{\frac{dt}{d{\tau}_t}}+
{\topstar{\Gamma}}_{{\beta}{\nu}}^{{\,}{\mu}}u_{(t)}^{\beta}
{\Big )}u_{(t)}^{\nu}=0.
\ena
The ${\mu}=0$ component of this equation (using a GTCS), in combination with 
equation (4), then immediately yields that $u_{(t)}^0$ must be constant along 
the particle's trajectory. Moreover, since $u_{(t)}^{\mu}u_{(t){\mu}}=-c^2$, we
also have $-(u_{(t)}^0)^2+u_{(t)}^iu_{(t)i}=-c^2$ which means that even
${\mid}{\vec {\bf u}}_t{\mid}{\equiv}{\sqrt{u_{(t)}^iu_{(t)i}}}$ is constant 
along the trajectory. Since ${\mid}{\vec {\bf u}}_t{\mid}$ is related to the
speed $w$ of the particle with respect to the FOs by the equation
${\mid}{\vec {\bf u}}_t{\mid}=w/{\sqrt{1-{\frac{w^2}{c^2}}}}$, we get the 
(maybe unexpected) result that {\em the cosmic expansion does not directly
affect speeds of material particles in quasi-metric relativity}. This means 
that the magnitude ${\mid}{\vec {\bf p}}_t{\mid}$ of a particle's 3-momentum is
not affected either, since the standard expression for the 4-momentum
$p_{(t)}^{\mu}=mu_{(t)}^{\mu}$ is valid. The difference of this result from its
counterpart in GR, where the magnitude of a freely-propagating particle's 
3-momentum changes with time in inverse proportionality to the scale factor, 
is very important.

On the other hand, it is straightforward to derive the usual expansion redshift
(and the corresponding time dilation) for a photon (or more generally, a null
particle) moving on a null geodesic using equations (14) (with ${\bar N}_t=1$) 
and (6) [2]. (Of course, this result is not in conflict with the fact that the 
speed $c$ of the photon with respect to the FOs is not affected by the cosmic 
expansion; in this respect the cosmic expansion does not distinguish between 
photons and material particles.) 

As we shall see, the fact that in quasi-metric cosmology, the momenta of 
photons are redshifted by the cosmic expansion but the momenta of material 
particles are not, is of crucial significance for the formulation of 
thermodynamics in quasi-metric space-time as presented in the next subsection.
\subsection{Equilibrium thermodynamics in quasi-metric 
space-time}
The fact that momenta of material particles, and thus their kinetic energies, 
are not affected by the cosmic expansion, implies that within the QMF, {\em the
cosmic expansion is not in general associated with work.} This means that in 
the QMF, the laws of thermodynamics will differ from their counterparts in GR. 
That is, since the cosmic expansion affects material particles differently, but
photons similarly in GR and in the QMF, the laws of thermodynamics must be 
dissimilar in the two theories.

Mathematically, the reason for this dissimilarity is that the volume element of
phase space, i.e., $dx^1dx^2dx^3dp_1dp_2dp_3$, will not be conserved along the
trajectories of material particles, but will change due to the cosmic expansion.
This must be so since the configuration comoving volume element $dx^1dx^2dx^3$ 
increases in time with a factor ${\frac{t^3}{t^3_0}}$, whereas the momentum 
comoving volume element $dp_1dp_2dp_3$ does not depend on $t$. This means that 
in the QMF, {\em Liouville's theorem does not hold for material particles}. 
Moreover, since the cosmic expansion does not directly affect the kinetic 
energy of material particles, the cosmic expansion does not directly affect gas
temperatures either, so that in the QMF, {\em the cosmic expansion does not 
directly cool a gas of material particles}. On the other hand, for photons, the
momentum volume element decreases in time with a factor ${\frac{t^3_0}{t^3}}$, 
compensating for the increase of the configuration volume element and ensuring 
that the volume element of phase space does not depend on the cosmic expansion.
This means that the cosmic expansion directly cools a photon gas, just as in 
GR. A third possibility is a gas of neutrinos where the lowest neutrino mass 
eigenstate (but not the two others) is null. Then only the null eigenstate will
be cooled directly by the cosmic expansion. However, in this paper we will 
assume that no neutrino mass eigenstate is null (see below for more comments).

Due to the violation of Liouville's theorem, expressions for the number density
$n_{{\rm p}_i}$, energy density ${\varepsilon}_{{\rm p}_i}{\equiv}
{\varrho}_{{\rm m}_i}c^2$ and pressure $p_{{\rm p}_i}$ for a noninteracting gas 
consisting of material particles of type ${\rm p}_i$ in kinetic equilibrium 
cannot be calculated unambiguously using standard quantum statistics. This is 
so since the effect of the cosmic expansion is to decrease said quantities but 
such that particle energies (and the temperature) remain constant. That is, 
(noninteracting) particle numbers and energies in any given comoving volume 
$V$ are constants, but arbitrary. In more detail, we note that the number 
${\Delta}N_{E_{{\rm p}_i}}$ of material particles with energy between 
$E_{{\rm p}_i}$ and $E_{{\rm p}_i}+{\Delta}E_{{\rm p}_i}$ is not affected by the 
cosmic expansion. Furthermore, let ${\Delta}g_{E_{{\rm p}_i}}$ denote the number 
of possible microstates a particle could occupy in one-particle phase space. 
Then we have ${\Delta}g_{E_{{\rm p}_i}}{\propto}V{\sqrt{E_{{\rm p}_i}^2-m_i^2c^4}}
E_{{\rm p}_i}dE_{{\rm p}_i}$ [4]. Note that ${\Delta}g_{E_{{\rm p}_i}}$ increases 
due to the cosmic expansion since $V{\propto}t^3$ and since particle energies 
$E_{{\rm p}_i}$ do not change. Next, the phase space occupancy function is given 
by $f(E_{{\rm p}_i},V){\equiv}{\frac{{\Delta}N_{E_{{\rm p}_i}}}{{\Delta}
g_{E_{{\rm p}_i}}}}{\propto}V^{-1}$ [4], so $f(E_{{\rm p}_i},V)$ will decrease due to
the cosmic expansion. This means that the number density [4] $n_{{\rm p}_i}
{\equiv}V^{-1} \sum_{E_{{\rm p}_i}}f(E_{{\rm p}_i},V){\Delta}g_{E_{{\rm p}_i}}{\propto}
V^{-1}$ will decrease due to the cosmic expansion. Similar results also apply 
to ${\varepsilon}_{{\rm p}_i}$ and $p_{{\rm p}_i}$. Note that
$f(E_{{\rm p}_i},V(t_2))={\frac{V(t_2)}{V(t_1)}}f(E_{{\rm p}_i},V(t_1))$; i.e.,
phase space occupancy functions can only be determined up to a volume factor.

The above reasoning implies that a knowledge of gas temperature and particle 
mass (and possibly chemical potential) is not sufficient to calculate 
$n_{{\rm p}_i}$, ${\varepsilon}_{{\rm p}_i}$ and $p_{{\rm p}_i}$; one must also know 
them at some arbitrary reference epoch $t_{\rm r}$, say. This means that,
using standard quantum statistics, it is only possible to calculate said 
quantities up to a multiplicative factor depending on cosmic scale, if such a 
gas is in free streaming and has never been in thermal equilibrium with 
photons. However, {\em ratios} of said quantities do not depend explicitly on
cosmic scale, and are given unambiguously from standard quantum statistics.
That is, to know said quantities at all epochs $t$, one must essentially 
{\em measure} one of them at epoch $t_{\rm r}$. As a consistency check of this 
result; for a gas consisting of noninteracting material particles the 
temperature $T(t)$ is constant so that the no-creation condition 
${\frac{{\partial}{\bar {\varepsilon}}}{{\partial}t}}=0$ together with equation
(16) yield
\eqa
n_{{\rm p}_i}(t)={\Big (}{\frac{t_{\rm r}}{t}}{\Big )}^3
n_{{\rm p}_i}(t_{\rm r}), 
\quad
{\varepsilon}_{{\rm p}_i}(t)={\Big (}{\frac{t_{\rm r}}{t}}{\Big )}^3
{\varepsilon}_{{\rm p}_i}(t_{\rm r}), \quad
p_{{\rm p}_i}(t)={\Big (}{\frac{t_{\rm r}}{t}}{\Big )}^3
p_{{\rm p}_i}(t_{\rm r}).
\ena
The ratios of these quantities follow from standard quantum statistics for the
case where $f(E_{{\rm p}_i},V)$ represents distribution functions valid for 
particles in a state of maximum entropy. These functions are the Bose-Einstein
and the Fermi-Dirac distributions for bosons and fermions, respectively (see, 
e.g., [4, 5] for derivation of explicit formulae), i.e.,
\eqa
{\frac{{\varepsilon}_{{\rm p}_i}}{n_{{\rm p}_i}}}=m_ic^2{\Big (}
\int_1^{\infty}{\frac{x^2{\sqrt{x^2-1}}dx}{{\exp}{\Big [}{\frac{m_ic^2x-
{\mu}_i}{k_{\rm B}T}}{\Big ]}{\pm}1}}{\Big )}{\times}
{\Big (} \int_1^{\infty}{\frac{x{\sqrt{x^2-1}}dx}{{\exp}{\Big [}
{\frac{m_ic^2x-{\mu}_i}{k_{\rm B}T}}{\Big ]}{\pm}1}}{\Big )}^{-1},
\ena
\eqa
{\frac{p_{{\rm p}_i}}{n_{{\rm p}_i}}}={\frac{1}{3}}m_ic^2{\Big (}
\int_1^{\infty}{\frac{(x^2-1)^{3/2}dx}
{{\exp}{\Big [}{\frac{m_ic^2x-{\mu}_i}{k_{\rm B}T}}{\Big ]}{\pm}1}}{\Big )}
{\times}{\Big (} \int_1^{\infty}{\frac{x{\sqrt{x^2-1}}dx}{{\exp}{\Big [}
{\frac{m_ic^2x-{\mu}_i}{k_{\rm B}T}}{\Big ]}{\pm}1}}{\Big )}^{-1},
\ena
where $k_{\rm B}$ is Boltzmann's constant, $m_i$ is particle mass and where 
${\mu}_i={\mu}_i(T)$ is chemical potential. The expressions for the similar 
ratios between the quantities $n_{{\rm {\bar p}}_i}$, 
${\varepsilon}_{{\rm {\bar p}}_i}$ and $p_{{\rm {\bar p}}_i}$ for the corresponding 
antiparticles can be found from equations (23) and (24) by letting 
${\mu}_i{\rightarrow}-{\mu}_i$. For neutrinos, in this paper we will assume 
that neutrino masses are small enough (but nonzero) so that they can be 
neglected together with neutrino chemical potentials. That is, for an 
extremely relativistic neutrino gas in kinetic equilibrium, standard quantum 
statistics yields [5]
\eqa
{\frac{{\varepsilon}_{{\nu}_i}}{n_{{\nu}_i}}}=
{\frac{7{\pi}^4}{180{\zeta}(3)}}k_{\rm B}T_{\nu}, \quad
p_{{\nu}_i}={\frac{1}{3}}{\varepsilon}_{{\nu}_i}, \quad
k_{\rm B}T_{\nu}{\gg}m_{\nu}c^2, \quad k_{\rm B}T_{\nu}{\gg}{\mu}_{\nu},
\ena
and identical formulae for the antineutrinos since neutrino chemical potentials
${\mu}_{{\nu}_i}$ have been neglected. We will set ${\mu}_{{\nu}_i}{\equiv}0$ for 
the rest of this paper.

Contrary to a gas of material particles, for a pure photon gas in kinetic
equilibrium, Liouville's theorem holds and the expressions for the number 
density $n_{\gamma}$ and energy density ${\varepsilon}_{\gamma}$ take their 
standard form [5]
\eqa
n_{\gamma}={\frac{2{\zeta}(3)k_{\rm B}^3T_{\gamma}^3}{{\pi}^2c^3{\hbar}^3}}, 
\qquad
{\varepsilon}_{\gamma}={\frac{{\pi}^2k_{\rm B}^4T_{\gamma}^4}{15c^3{\hbar}^3}}, 
\qquad
p_{\gamma}={\frac{{\varepsilon}_{\gamma}}{3}}.
\ena 
In equation (26), $n_{\gamma}$ and ${\varepsilon}_{\gamma}$ depend on time via 
the temperature $T_{\gamma}(t)={\frac{t_{\rm r}}{t}}T_{\gamma}(t_{\rm r})$,
i.e., $T_{\gamma}(t)$ changes as the inverse of the scale factor, just as in 
GR. This means that the cosmic expansion changes $n_{\gamma}$ and 
${\varepsilon}_{\gamma}$ unambiguously via $T_{\gamma}(t)$. This is also 
consistent with equation (16) and the no-creation condition.

Next we notice that it is not sufficient to consider a pure photon gas for 
applications to the early universe, since one must deal with a plasma 
consisting of photons interacting with material (elementary) particles. 
Therefore, one must consider a dilute, weakly interacting gas consisting of 
both photons and material particles (except possibly the neutrinos) satisfying 
the conditions of thermal and chemical equilibrium to a good approximation. On 
the other hand, depending on the plasma temperature, neutrinos/antineutrinos 
may or may not be in thermal equilibrium with the plasma. For such a gas, 
equations (26) can still be used for the photons even when the plasma 
temperature $T(t)$ does not vary as the inverse of the scale factor. This is 
possible due to the fact that the plasma contains traces of protons, so that 
inelastic scattering reactions $e^{\pm}+p{\rightarrow}e^{\pm}+p+{\gamma}$ occur 
rapidly and are capable of producing a sufficient number of photons for 
equation (26) to hold during the whole relevant temperature range. Thus neither
photon number nor total particle number will in general be conserved, which is 
necessary to maintain the relevant particle species in thermal equilibrium.

For the cosmic plasma, the cosmic expansion will directly cool the photons, 
while the material particles are cooled indirectly via thermal contact with the 
photons. So the effective heat capacity of the plasma is higher than for a pure
photon gas and $T(t)$ decreases more slowly. This results in an effective 
``time-averaged'' violation of Liouville's theorem for the photons as well, 
since heating from the material particles implies that the decrease of momentum
volume will no longer exactly cancel the increase in configuration volume, so 
that the volume of phase space increases with time. Similarly, the cooling of 
the material particles means that averaged over time, some of the increase in 
configuration volume is canceled by a decrease in momentum volume, so that the 
increase in phase space volume with time will be smaller than if the gas of 
material particles were not interacting with the photons. Notice that, for a 
plasma in thermal equilibrium, the increase in phase space volume must of 
course be identical for the photons and the material particles. (This also 
applies to the neutrinos if they are in thermal equilibrium with the plasma.)

The above discussion implies that, for a gas of material particles in thermal 
equilibrium with photons and possibly neutrinos, as long as said photon-number
violating reactions occur rapidly, it is possible to directly calculate 
$n_{\gamma}$, ${\varepsilon}_{\gamma}$, $n_{{\rm p}_i}$, ${\varepsilon}_{{\rm p}_i}$ 
and $p_{{\rm p}_i}$ (or $n_{{\rm {\bar p}}_i}$, ${\varepsilon}_{{\rm {\bar p}}_i}$ and 
$p_{{\rm {\bar p}}_i}$ for the corresponding antiparticles) unambiguously using 
standard expressions from quantum statistics. That is, thermal equilibrium 
with the photons determines the ratios
\eqa
{\frac{n_{{\rm p}_i}}{n_{\gamma}}}=
{\frac{m^3_ic^6g_i}{4{\zeta}(3)(k_{\rm B}T)^3}} \int_1^{\infty}
{\frac{x{\sqrt{x^2-1}}dx}
{{\exp}{\Big [}{\frac{m_ic^2x-{\mu}_i}{k_{\rm B}T}}{\Big ]}{\pm}1}},
\ena
\eqa
{\frac{{\varepsilon}_{{\rm p}_i}}{n_{\gamma}}}=
{\frac{m^4_ic^8g_i}{4{\zeta}(3)(k_{\rm B}T)^3}}
\int_1^{\infty}{\frac{x^2{\sqrt{x^2-1}}dx}
{{\exp}{\Big [}{\frac{m_ic^2x-{\mu}_i}{k_{\rm B}T}}{\Big ]}{\pm}1}}, 
\ena
for particle type ${\rm p}_i$ with mass $m_i$, chemical potential ${\mu}_i$ and
$g_i$ internal degrees of freedom, plus similar ratios involving the pressure. 
Note that similar ratios involving the neutrinos only hold if the neutrinos 
are in thermal equilibrium with the plasma and that this is possible only if
neutrino-number violating reactions occur rapidly. In combination with equation
(26), the ratios (27) and (28) yield explicit formulae for $n_{{\rm p}_i}$ and 
${\varepsilon}_{{\rm p}_i}$, the same as valid for standard cosmology. Also note 
the formula, valid for nonrelativistic fermion species in thermal 
equilibrium with the plasma, [5]
\eqa
n_{{\rm p}_i}(t)-n_{{\rm {\bar p}}_i}(t){\approx}
{\frac{2g_i}{c^3{\hbar}^3}}{\Big (}{\frac{m_ic^2k_{\rm B}T}{2{\pi}}}{\Big )}
^{3/2}{\sinh}{\Big (}{\frac{{\mu}_i}{k_{\rm B}T}}{\Big )}
{\exp}{\Big (}-{\frac{m_ic^2}{k_{\rm B}T}}{\Big )}, \qquad 
k_{\rm B}T{\ll}m_{{\rm p}_i}c^2.
\ena
Furthermore, the no-creation condition applied to all particles yields the 
first law of thermodynamics. This may be written in the form $d[{\varepsilon}V]
=-p_{\gamma}dV+{\mu}d[nV]$ (in shorthand notation), where $V$ is a
fiducial comoving volume and where $p_{\gamma}$ is the pressure associated with
the photons. The first law takes this form because the cosmic expansion affects
photons and material particles differently so that only the photons may be 
considered doing work. Written out as a sum over all particles species, 
the first law of thermodynamics then takes the form
\eqa
\sum_i [{\dot {\varepsilon}}_{{\rm p}_i}(t)+{\dot {\varepsilon}}_
{{\rm {\bar p}}_i}(t)+2{\dot {\varepsilon}}_{{\nu}_i}(t)]+
{\dot {\varepsilon}}_{\gamma}(t)=-{\frac{1}{t}} \sum_i
[3{\varepsilon}_{{\rm p}_i}(t)+3{\varepsilon}_{{\rm {\bar p}}_i}(t)
+6{\varepsilon}_{{\nu}_i}(t)]-{\frac{4}{t}}{\varepsilon}_{\gamma}(t)
\nonumber \\
+\sum_i {\mu}_i(T){\Big [}{\dot n}_{{\rm p}_i}(t)-{\dot n}_{{\rm {\bar p}}_i}
(t)+{\frac{3}{t}}[n_{{\rm p}_i}(t)-n_{{\rm {\bar p}}_i}(t)]{\Big ]},
\ena
where a ``dot'' denotes a time derivative, and where the sum runs over all 
particle species ${\rm p}_i$ in thermal equilibrium. The counterpart to 
equation (30) for the pressures can be found directly from the 
quantum-statistical expressions for $n$, ${\varepsilon}$ and $p$ by taking 
differentials using integration by parts. The result may be written in the 
form $dp={\frac{{\varepsilon}+p-{\mu}n}{T}}dT+nd{\mu}$ [5]. Written
out as a sum over all particle species this yields
\eqa
 \sum_i [{\dot p}_{{\rm p}_i}(t)+{\dot p}_{{\rm {\bar p}}_i}(t)+
2{\dot p}_{{\nu}_i}(t)]+{\dot p}_{\gamma}(t) 
= \sum_i {\dot {\mu}}_i(T)[n_{{\rm p}_i}(t)-n_{{\rm {\bar p}}_i}(t)]
+{\frac{{\dot T}}{T}}{\Big (}{\varepsilon}_{\gamma}(t)+p_{\gamma}(t)
\nonumber \\
+ \sum_i {\Big [}
{\varepsilon}_{{\rm p}_i}(t)+{\varepsilon}_{{\rm {\bar p}}_i}(t)
+2{\varepsilon}_{{\nu}_i}(t)+p_{{\rm p}_i}(t)+p_{{\rm {\bar p}}_i}(t)
+2p_{{\nu}_i}(t)-{\mu}_i(T)[n_{{\rm p}_i}(t)-n_{{\rm {\bar p}}_i}(t)]
{\Big ]}{\Big )}.
\ena
From the first law of thermodynamics (30) and the explicit expressions for
$n_{{\rm p}_i}$ and ${\varepsilon}_{{\rm p}_i}$ obtained from equations (26), (27) 
and (28), we are able to find an expression for the temperature evolution of 
the cosmic plasma (if no decoupling occurs). We find
\eqa
{\frac{{\dot T}(t)}{T(t)}}=-{\frac{{\beta}(T)}{t}}, \qquad
{\beta}(T){\equiv}{\frac{{\varepsilon}_{\gamma}+{\frac{3}{4}} \sum_i[
2{\varepsilon}_{{\nu}_i}+{\varepsilon}_{{\rm p}_i}+
{\varepsilon}_{{\rm {\bar p}}_i}-{\mu}_i(n_{{\rm p}_i}-n_{{\rm {\bar p}}_i})]}
{{\varepsilon}_{\gamma}+ \sum_j {\{}2{\varepsilon}_{{\nu}_j}+
{\varepsilon}_{{\rm p}_j}+{\varepsilon}_{{\rm {\bar p}}_j}
+{\frac{m_jc^2}{4}}(I_{{\rm p}_j}+I_{{\rm {\bar p}}_j})+Z_j({\mu}_j){\}}}},
\ena
\eqa
Z_j({\mu}_j){\equiv}{\frac{1}{4}}{\Big (}T{\frac{d{\mu}_j}{dT}}-2{\mu}_j
{\Big )}{\Big (}3n_{{\rm p}_j}-3n_{{\rm {\bar p}}_j}+
J_{{\rm p}_j}-J_{{\rm {\bar p}}_j}{\Big )} \nonumber \\ 
-{\frac{{\mu}_j}{4m_jc^2}}{\Big (}T{\frac{d{\mu}_j}{dT}}-{\mu}_j{\Big )}
{\Big (}K_{{\rm p}_j}+K_{{\rm {\bar p}}_j}
+2L_{{\rm p}_j}+2L_{{\rm {\bar p}}_j}{\Big )},
\ena
\eqa
I_{{\rm p}_j}{\equiv}C_j
\int_1^{\infty}{\frac{x^2dx}{{\sqrt{x^2-1}}{\Big (}
{\exp}{\Big [}{\frac{m_jc^2x-{\mu}_j}{k_{\rm B}T}}{\Big ]}{\pm}1{\Big )}}}, 
\qquad C_j{\equiv}{\frac{m^3_jc^3g_j}{2{\pi}^2{\hbar}^3}},
\ena
\eqa
J_{{\rm p}_j}{\equiv}C_j
\int_1^{\infty}{\frac{xdx}{{\sqrt{x^2-1}}{\Big (}
{\exp}{\Big [}{\frac{m_jc^2x-{\mu}_j}{k_{\rm B}T}}{\Big ]}{\pm}1{\Big )}}},
\ena
\eqa
K_{{\rm p}_j}{\equiv}C_j
\int_1^{\infty}{\frac{dx}{{\sqrt{x^2-1}}{\Big (}
{\exp}{\Big [}{\frac{m_jc^2x-{\mu}_j}{k_{\rm B}T}}{\Big ]}{\pm}1{\Big )}}},
\quad L_{{\rm p}_j}{\equiv}C_j
\int_1^{\infty}{\frac{{\sqrt{x^2-1}}dx}{
{\exp}{\Big [}{\frac{m_jc^2x-{\mu}_j}{k_{\rm B}T}}{\Big ]}{\pm}1}}.
\ena
We see that as long as a significant number of material particles are in 
thermal equilibrium with the photons, the temperature will drop more slowly
than if (almost) only photons were present. As mentioned earlier, we have 
assumed that no neutrino mass eigenstate is null in the above formulae. 
However, from experiments it is not ruled out that the least massive neutrino 
mass eigenstate is indeed null. For this case the above formulae must be 
modified by summing over neutrino mass eigenstates rather than flavour 
eigenstates, taking into account that the cosmic expansion will redshift the 
energy of the null mass eigenstate. Then it will have the same status as the 
photons in the above formulae.

As we have seen above, the first law of thermodynamics in its standard form 
does not apply to quasi-metric cosmology. On the other hand, the second law of 
thermodynamics does apply in its standard form. This follows from the fact that
it is possible to derive [4] the usual expression for the entropy (for a closed
system) directly from the general definition $S{\equiv}-k_{\rm B} \sum_{\alpha}
P_{\alpha}{\ln}P_{\alpha}$ in terms of the probability $P_{\alpha}$ that the system 
is in microstate $\alpha$, without using the first law. This derivation is 
valid for quasi-metric cosmology as well. Thus one may define {\em the entropy 
density} $s(t)$ for the cosmic plasma just as in standard cosmology, i.e.,
\eqa
s(t){\equiv}{\frac{1}{T}}{\Big (} \sum_i{\Big [}{\varepsilon}_{{\rm p}_i}
+{\varepsilon}_{{\rm {\bar p}}_i}+p_{{\rm p}_i}+p_{{\rm {\bar p}}_i}-
{\mu}_i(n_{{\rm p}_i}-n_{{\rm {\bar p}}_i})+2{\varepsilon}_{{\nu}_i}
+2p_{{\nu}_i}{\Big ]}+{\frac{4}{3}}{\varepsilon}_{\gamma}{\Big )},
\ena
since we have assumed that ${\mu}_{{\nu}_i}{\equiv}0$. By inserting equations 
(22) and (26) into equation (37) and multiplying with a fiducial comoving 
volume $V$, it is straightforward to see that the entropy $S{\equiv}s(t)V=
{\frac{[{\varepsilon}+p-{\mu}n]V}{T}}$ does not depend on the scale factor, 
neither for a gas consisting of non-interacting material particles only, nor 
for a pure photon gas (for which $\mu =0$). This is as expected, since for 
the first case $T$ is constant and for the second case Liouville's theorem 
holds, so the cosmic expansion may be associated with work. However, for a gas 
consisting of both photons and material particles in thermal equilibrium, only 
the partial pressure of the photons may be considered doing work and $T^3V$ 
will not be constant. Moreover, there will be a net heat transfer from the
material particles to the plasma. This heat transfer may be treated as if
the plasma were heated by an external source and its magnitude is precisely
the work $\sum_i (p_{{\rm p}_i}+p_{{\rm {\bar p}}_i})dV$ that should have been done 
by the material particles according to standard cosmology. Taking into account 
both the pressure work done by the photons and said net heat transfer, the 
second law of thermodynamics then takes its standard form.

The above considerations would indicate that for a general plasma, the total 
entropy in a comoving volume should not be conserved. To see this explicitly; 
from the second law of thermodynamics we find 
$TdS=d[{\varepsilon}V]+pdV-{\mu}d[nV]=(p-p_{\gamma})dV$, where we have used the 
first law of thermodynamics in the last step. This means that $s(t)$ will in 
general decrease more slowly than $t^{-3}$, since we find that
\eqa
{\frac{d}{dt}}s(t)=-{\frac{3}{t}}{\Big [}s(t)-{\frac{1}{T(t)}} 
\sum_i [{p_{{\rm p}_i}}(t)+p_{{\rm {\bar p}}_i}(t)+2p_{{\nu}_i}(t)]{\Big ]}.
\ena
We see from equation (38) that if a significant number of relativistic
material particles is in thermal equilibrium with the photons, entropy will
increase with time. In particular this applies to neutrinos before they 
decouple since the neutrinos are ultrarelativistic. On the other hand, after
neutrino decoupling, only nonrelativistic particle species are in thermal
equilibrium with the photons and $s(t)V$ will be constant to high accuracy.
\subsection{Thermodynamics during neutrino decoupling}
In the previous section we have assumed that all relevant interparticle 
reaction rates are much faster than the fractional temperature change of the
cosmic plasma due to the cooling effect of the cosmic expansion. This 
assumption is necessary in order to treat the plasma as (nearly) in a state of 
thermodynamical equilibrium. For example, to keep neutrinos of all flavours
in thermal equilibrium, it is necessary to increase the number of neutrinos as 
the temperature decreases. For temperatures where muons can be neglected, this 
can be achieved if the reactions
\eqa
&&{\gamma}+{\gamma}{\ }{\rightleftharpoons}{\ }e^++e^-, \qquad
e^++e^-{\ }{\rightleftharpoons}{\ }{\bar {\nu}}_i+{\nu}_i, \nonumber \\
&&{\nu}_i+{\bar {\nu}}_i{\ }{\rightarrow}{\ }{\nu}_j+{\bar {\nu}}_j, \qquad
{\gamma}+e^{\pm}{\ }{\rightarrow}{\ }e^{\pm}+{\bar {\nu}}_i+{\nu}_i, \qquad
i,j{\in}{\{ }e,{\mu},{\tau}{\} },
\ena
proceed sufficiently fast. (Here, the first and second reactions represent
annihilation/pair creation balances and the fourth reaction is called the 
photoneutrino process. The annihilation reaction is more effective than the 
photoneutrino process in producing new neutrinos for the relevant temperature 
range.) Besides, elastic scattering processes between neutrinos and electrons 
or positrons are required to be sufficiently efficient so that the electrons 
and positrons will act as heat conductors responsible for the thermal contact 
between the photons and the neutrinos in thermal equilibrium. These scattering 
processes are given by
\eqa
e^{\pm}+{\nu}_i{\ }{\rightarrow}{\ }e^{\pm}+{\nu}_i, \qquad
e^{\pm}+{\bar {\nu}}_i{\ }{\rightarrow}{\ }e^{\pm}+{\bar {\nu}}_i.
\ena
However, for sufficiently low temperatures, weak interaction reaction rates 
will not be fast compared to the cooling rate of the plasma, leading to 
neutrino decoupling. (On the other hand, the first reaction shown in equation 
(39) is electromagnetic and its rate is rapid enough to maintain the 
equilibrium number of electrons/positrons for the relevant temperature range.)

Since neutrinos are not in thermodynamic equilibrium during neutrino 
decoupling, one must in principle solve the Boltzmann equation to find the
neutrino distribution functions (see appendix B). But since approximate 
forms of said distribution functions should be sufficient for mere estimates 
of the relevant thermodynamic quantities, we shall rather set up an 
approximate model for thermodynamics during neutrino decoupling (valid for 
quasi-metric cosmology). In this approximate model, neutrino decoupling 
proceeds in two steps. The first, initial stage occurs when the scattering 
processes (40) become too slow to maintain thermodynamical equilibrium for a 
non-negligible part of the low-energy neutrinos. That is, for neutrinos of type
${\nu}_i$ with energy $E_{{\nu}_i}$ below an energy threshold ${E_{\rm D}}_i(T)$, 
said scattering reactions are so slow that these neutrinos will in effect be 
decoupled from the plasma. (Of course this is not absolute since there is a 
small chance that any low-energy neutrino may interact with a sufficiently 
high-energy electron/positron. This means that the values of ${E_{\rm D}}_i(T)$ 
are not sharp and should be taken as estimated values.) Since each 
${E_{\rm D}}_i(T)$ increases as the temperature decreases, the first stage of 
neutrino decoupling begins at the low end of the neutrino energy spectrum 
and proceeds to higher energies. For a while, the second and fourth reactions 
shown in equation (39) proceed fast enough to supply a sufficient number of 
new neutrinos so that neutrinos with energy $E_{{\nu}_i}>{E_{\rm D}}_i(T)$ can 
still be considered to be in thermodynamical equilibrium in spite of the fact 
that the system "leaks". (Since the cosmic plasma loses a matter component as 
if it were an open system leaking matter into the surroundings, the cosmic 
plasma must be treated effectively as an open system during neutrino 
decoupling.) However, when the temperature drops below a critical level 
$T_{\rm C}$, said reactions (39) become too slow to produce a sufficient number 
of new neutrinos necessary to maintain thermodynamical equilibrium for 
neutrinos with $E_{{\nu}_i}>{E_{\rm D}}_i(T)$, initiating the second stage of 
neutrino decoupling. This may happen even when most neutrinos are still in 
thermal contact with the plasma via the scattering reactions shown in equation 
(39). Note that the two stages of neutrino decoupling may begin at epochs well 
separated in time (and temperature).

After the second stage of neutrino decoupling has begun, neutrino number 
densities will continue to drop compared to equilibrium values. The final 
energy distribution of the decoupled neutrinos will depend on how 
fast each ${E_{\rm D}}_i(T)$ increases compared to scattering rates of the 
neutrinos still in thermal contact with the plasma. If said scattering rates 
are much faster than the increase of each ${E_{\rm D}}_i(T)$, most neutrinos 
will decouple via interactions by falling below ${E_{\rm D}}_i(T)$. On the other
hand, if each ${E_{\rm D}}_i(T)$ increases much faster than said scattering 
rates, the energy distribution of the highest energy neutrinos will be 
``frozen in''. However, at later times, where ${E_{\rm D}}_i(T)$ is so high
that a significant fraction of a neutrino equilibrium energy distribution with 
temperature $T$ would fall below ${E_{\rm D}}_i(T)$, the model is expected to 
fail since there is an insufficient number of low-energy neutrinos available 
and these interact too slowly to maintain a high-energy thermal tail. On the 
other hand, the high-energy neutrinos interact much more rapidly and tend to 
end up as low-energy neutrinos for every interaction. This means that according
to quasi-metric theory, one would expect that the high-energy part of the relic 
neutrino background should be depleted and thus deviate 
significantly from that of thermal neutrino distributions.

Since the assumption of thermodynamical equilibrium does not hold for the 
neutrinos during neutrino decoupling, some of the formulae found in the 
previous section must be modified. To do that, we first split up total 
neutrino number and energy densities into effective and decoupled parts, i.e., 
$n_{{\nu}_i}(t)=n^{\rm eff}_{{\nu}_i}(t)+n^{\rm dec}_{{\nu}_i}(t)$ and
${\varepsilon}_{{\nu}_i}(t)={\varepsilon}^{\rm eff}_{{\nu}_i}(t)+
{\varepsilon}^{\rm dec}_{{\nu}_i}(t)$. We now assume that the second stage of 
neutrino decoupling starts at epoch $t_{\rm C}$ and the ``critical'' temperature 
$T_{\rm C}$. Moreover, we assume that all neutrinos of type ${\nu}_i$ with 
energies above the threshold energy ${E_{\rm D}}_i(T)$ are effectively in 
thermal contact with the photon plasma and approximately follow a (modified) 
Fermi-Dirac distribution with temperature equal to that of the photon plasma. 
Neutrinos with energy below ${E_{\rm D}}_i(T)$ are decoupled and their phase 
space distribution function is unknown. By assumption we thus have the 
approximate model {\em ansatz}
\eqa
n^{\rm eff}_{{\nu}_i}{\approx}{\frac{h_i(t)k_{\rm B}^3T^3}{2{\pi}^2
{\hbar}^3c^3}} \int_{{\frac{{E_{\rm D}}_i}{k_{\rm B}T}}}^{\infty}
{\frac{x^2dx}{1+{\exp}(x)}}, \quad
{\varepsilon}^{\rm eff}_{{\nu}_i}{\approx}{\frac{h_i(t)k_{\rm B}^4T^4}{2{\pi}^2
{\hbar}^3c^3}} \int_{{\frac{{E_{\rm D}}_i}{k_{\rm B}T}}}^{\infty}
{\frac{x^3dx}{1+{\exp}(x)}}, 
\ena
where $h_i(t)$ are ``scaling'' functions taking into account the fact that 
neutrino number densities may be smaller than equilibrium values even when 
there is good thermal contact with the cosmic plasma. As mentioned above,
equation (41) is expected to fail for ${E_{\rm D}}_i(T){\sim}k_{\rm B}T$, but
for the lack of alternatives we will assume that it is approximately valid 
througout neutrino decoupling.

Now, since equation (41) only represents an approximate model, a 
simplification of it can be justified if the results are approximately 
unchanged. Therefore, we will assume that neutrino effective number and energy 
densities do not depend significantly on neutrino type, so that 
$n^{\rm eff}_{{\nu}_i}{\approx}n^{\rm eff}_{\nu}$ and
${\varepsilon}^{\rm eff}_{{\nu}_i}{\approx}{\varepsilon}^{\rm eff}_{\nu}$, where
\eqa
n^{\rm eff}_{{\nu}}{\equiv}{\frac{h(t)k_{\rm B}^3T^3}{2{\pi}^2
{\hbar}^3c^3}} \int_{{\frac{E_{\rm D}}{k_{\rm B}T}}}^{\infty}
{\frac{x^2dx}{1+{\exp}(x)}}, \quad
{\varepsilon}^{\rm eff}_{{\nu}}{\equiv}{\frac{h(t)k_{\rm B}^4T^4}{2{\pi}^2
{\hbar}^3c^3}} \int_{{\frac{E_{\rm D}}{k_{\rm B}T}}}^{\infty}
{\frac{x^3dx}{1+{\exp}(x)}}. 
\ena
In equation (42), the functions $h(t)$ and $E_{\rm D}$ are expected to 
sufficiently approximate the functions $h_i(t)$ and ${E_{\rm D}}_i$, 
respectively.

For $t<t_{\rm C}$ thermal equilibrium holds to good approximation for 
$E_{\nu}>E_{\rm D}$ so $h(t)=1$. For $t>t_{\rm C}$ equation (42) yields 
\eqa
{\dot n}^{\rm eff}_{{\nu}}(t)={\Big {\{}}
3{\frac{{\dot T}}{T}}+{\frac{{\dot h}}{h}}{\Big {\}}}n^{\rm eff}_{{\nu}}(t)
+{\frac{h(t)E_{\rm D}^2}{2{\pi}^2{\hbar}^3c^3}}{\Big [}
{\frac{E_{\rm D}-T{\frac{dE_{\rm D}}{dT}}}{{\exp}
({\frac{E_{\rm D}}{k_{\rm B}T}})+1}}{\Big ]}{\frac{{\dot T}}{T}} \nonumber \\
{\approx}-{\Big {\{}}{\frac{3}{t}}+{\frac{2}{3{\zeta}(3)}}
 \int_0^{{\frac{E_{\rm D}}{k_{\rm B}T}}}{\frac{x^2dx}{1+{\exp}(x)}}
{\Gamma}^{\rm eff}_{\rm scat}-{\Gamma}_{\rm ann}^{\rm net}
{\Big {\}}}n^{\rm eff}_{{\nu}}(t)
+{\frac{h(t)E_{\rm D}^2}{2{\pi}^2{\hbar}^3c^3}}{\Big [}
{\frac{E_{\rm D}-T{\frac{dE_{\rm D}}{dT}}}{{\exp}
({\frac{E_{\rm D}}{k_{\rm B}T}})+1}}{\Big ]}{\frac{{\dot T}}{T}},
\ena
where ${\Gamma}^{\rm eff}_{\rm scat}$ (see equation (49) below) is the thermally 
averaged effective interaction rate per particle of the scattering reactions 
shown in equation (40), and where ${\Gamma}_{\rm ann}^{\rm net}$ is the average net 
annihilation rate of the annihilation processes (producing neutrinos) shown in 
equation (39). Note that the second term of the second line of equation (43) 
represents the number density decoupling rate via said scattering processes, 
while the last term represents the number density decoupling rate due to the 
change with time of the factor ${\frac{E_{\rm D}}{k_{\rm B}T}}$. These two terms
represent the transfer rate of neutrinos to the decoupled part, so for each 
neutrino type we must have ($n^{\rm dec}_{{\nu}_i}{\approx}n^{\rm dec}_{{\nu}}$)
\eqa
{\dot n}^{\rm dec}_{{\nu}}(t){\approx}-{\frac{3}{t}}n^{\rm dec}_{{\nu}}(t)
+{\frac{2}{3{\zeta}(3)}} \int_0^{{\frac{E_{\rm D}}{k_{\rm B}T}}}{\frac{x^2dx}
{1+{\exp}(x)}}{\Gamma}^{\rm eff}_{\rm scat}n^{\rm eff}_{{\nu}}(t)
-{\frac{h(t)E_{\rm D}^2}{2{\pi}^2{\hbar}^3c^3}}{\Big [}
{\frac{E_{\rm D}-T{\frac{dE_{\rm D}}{dT}}}{{\exp}
({\frac{E_{\rm D}}{k_{\rm B}T}})+1}}{\Big ]}{\frac{{\dot T}}{T}},
\ena
for the number density of decoupled neutrinos (of each type).
Moreover, equation (43) yields
\eqa
{\frac{{\dot h}}{h}}{\approx}-{\frac{3}{t}}(1-{\beta}(T))-
{\frac{2}{3{\zeta}(3)}} \int_0^{{\frac{E_{\rm D}}{k_{\rm B}T}}}{\frac{x^2dx}
{1+{\exp}(x)}}{\Gamma}^{\rm eff}_{\rm scat}+
{\Gamma}_{\rm ann}^{\rm net}.
\ena
Note that, for $t<t_{\rm C}$, $T>T_{\rm C}$, the left hand side of equation (45) 
vanishes, determining ${\Gamma}_{\rm ann}^{\rm net}$ for this case.

We will now assume that the annihilation neutrinos thermalize quickly via the
scattering reactions shown in equation (40), even for $t>t_{\rm C}$. From 
equations (42) and (45) it then follows that
\eqa
{\dot {\varepsilon}}^{\rm eff}_{{\nu}}(t){\approx}{\Big {\{}}
{\frac{{\dot T}}{T}}-{\frac{3}{t}}+{\Gamma}_{\rm ann}^{\rm net}-
{\frac{2}{3{\zeta}(3)}} \int_0^{{\frac{E_{\rm D}}{k_{\rm B}T}}}{\frac{x^2dx}
{1+{\exp}(x)}}{\Gamma}^{\rm eff}_{\rm scat}
{\Big {\}}}{\varepsilon}^{\rm eff}_{{\nu}}(t)
\nonumber \\
+{\frac{h(t)E_{\rm D}^3}{2{\pi}^2{\hbar}^3c^3}}{\Big [}
{\frac{E_{\rm D}-T{\frac{dE_{\rm D}}{dT}}}{{\exp}
({\frac{E_{\rm D}}{k_{\rm B}T}})+1}}{\Big ]}{\frac{{\dot T}}{T}}.
\ena
The rate of decoupled energy density is found approximately from equation
(46) by taking into account the average energy of the decoupled 
neutrinos, i.e., (${\varepsilon}^{\rm dec}_{{\nu}_i}{\approx}
{\varepsilon}^{\rm dec}_{{\nu}}$)
\eqa
{\dot {\varepsilon}}^{\rm dec}_{{\nu}}(t){\approx}-{\frac{3}{t}}
{\varepsilon}^{\rm dec}_{{\nu}}(t)
+{\frac{2k_{\rm B}T}{3{\zeta}(3)}} 
\int_0^{{\frac{E_{\rm D}}{k_{\rm B}T}}}{\frac{x^3dx}
{1+{\exp}(x)}}{\Gamma}^{\rm eff}_{\rm scat}n^{\rm eff}_{{\nu}}(t)
-{\frac{h(t)E_{\rm D}^3}{2{\pi}^2{\hbar}^3c^3}}{\Big [}
{\frac{E_{\rm D}-T{\frac{dE_{\rm D}}{dT}}}{{\exp}
({\frac{E_{\rm D}}{k_{\rm B}T}})+1}}{\Big ]}{\frac{{\dot T}}{T}}.
\ena
Note that, if the second term on the right hand side of equation (47) is much 
larger than the last term, decoupling via particle interactions dominates the 
energy transfer to the decoupled neutrinos, while if it is the other way 
around, the neutrinos effectively ``freeze out''. In the second case, if 
$E_{\rm D}(t)$ increases much faster than $t$, the end of the second stage of
neutrino decoupling may be treated as an instantaneous process.

We can use equations (46) and (47) in combination with equation (30) to
find the value of ${\beta}(T)$ during neutrino decoupling. We find
\eqa
{\beta}(T){\approx}{\frac{ {\varepsilon}_{\gamma}+{\frac{3}{4}}
 \sum_i[{\varepsilon}_{{\rm p}_i}+{\varepsilon}_{{\rm {\bar p}}_i}
-{\mu}_i(n_{{\rm p}_i}-n_{{\rm {\bar p}}_i})]}
{{\varepsilon}_{\gamma}+{\frac{3}{2}}{\varepsilon}^{\rm eff}_{{\nu}}+
\sum_j {\{}{\varepsilon}_{{\rm p}_j}+
{\varepsilon}_{{\rm {\bar p}}_j}+{\frac{m_jc^2}{4}}(I_{{\rm p}_j}+
I_{{\rm {\bar p}}_j})+Z_j({\mu}_j){\}}}} \nonumber \\
+{\frac{
t{\Gamma}_{\rm ann}^{\rm net}{\frac{3}{2}}{\varepsilon}^{\rm eff}_{{\nu}}-
{\frac{t}{{\zeta}(3)}}{\Gamma}_{\rm scat}^{\rm eff}[ \int_0^
{\frac{E_{\rm D}}{k_{\rm B}T}}{\frac{x^2dx}{1+e^x}}{\varepsilon}_{\nu}^
{\rm eff}- \int_0^{\frac{E_{\rm D}}{k_{\rm B}T}}{\frac{x^3dx}{1+e^x}}
k_{\rm B}Tn_{\nu}^{\rm eff}]}
{{\varepsilon}_{\gamma}+{\frac{3}{2}}{\varepsilon}^{\rm eff}_{{\nu}}+
\sum_j {\{}{\varepsilon}_{{\rm p}_j}+
{\varepsilon}_{{\rm {\bar p}}_j}+{\frac{m_jc^2}{4}}(I_{{\rm p}_j}+
I_{{\rm {\bar p}}_j})+Z_j({\mu}_j){\}}}}.
\ena
Note that equation (48) is valid independent of the particular form the
energy transfer term takes in equations (46) and (47). Also note that when the 
first stage of neutrino decoupling starts, ${\beta}(T)$ will increase
relative to its equilibrium value at the same temperature since part of the
neutrino energy spectrum will be decoupled (combine equations (45) and (48) to 
see this explicitly). Thus the production of new neutrinos to counteract the
decoupling rate will in effect {\em decrease} the heat capacity of the
cosmic plasma. However, when the second stage of neutrino decoupling begins,
the value of ${\beta}(T)$ is expected to drop (suddenly) when the production
of new neutrinos effectively ceases. This behaviour can be understood as a 
consequence of the fact that the energy consumed by production of new neutrinos
will drop to almost zero, while the decoupling neutrinos furnish a net thermal
energy transfer to the plasma, so that the heat capacity of the plasma will 
{\em increase}. Later, ${\beta}(T)$ will again increase as more and more of the 
neutrinos become thermally disconnected from the cosmic plasma.
\section{Neutrino decoupling}
\subsection{General remarks}
The thermal history of the quasi-metric universe is defined by the temperature
evolution $T(t)$ of the cosmic radiation background. We see from equations 
(32) and (48) that $T(t)$ depends on the function ${\beta}(T)$, again 
depending on the number of neutrinos and material particle species present in 
thermal equilibrium with the photons. In particular, as estimated in [4], after
the era of electron-positron annihilation ended at a temperature of 
$\sim$$2{\times}$10$^8$ K ($\sim$20 keV$/k_{\rm B}$) (assuming an exess of 
electrons over positrons $(n_{e^-}-n_{e^+})/s$ of order $10^{-9}/k_{\rm B}$), in the
SBB scenario neutrinos were no longer in thermal contact with the photons, and 
the plasma consisted of a nearly pure photon gas (with traces of protons and 
electrons) so that ${\beta}(T) \approx 1$. Note that after neutrino decoupling,
equations (29) (with ${\mu}_e{\neq}0$) and (38) imply that $(n_{e^-}-n_{e^+})/s$ 
is approximately constant in the relevant temperature range so that the 
estimate made in [4] to find at which temperature the number of positrons will 
be negligible, is valid for quasi-metric cosmology as well. (Alternatively, 
assuming a baryon to photon number $n_b/n_{\gamma}{\approx}
(n_{e^-}-n_{e^+})/n_{\gamma}{\approx}6{\times}10^{-10}$, we find from equation (29) 
that the fraction $n_{e^+}/n_{e^-}={\exp}(-2{\mu}_e/k_{\rm B}T)$ becomes negligible
(of order $0.001-0.01$) for temperatures of about $\sim$17-18 keV$/k_{\rm B}$. 
This is consistent with said estimate made in [4].) Moreover, the age of the 
quasi-metric universe at the end of electron-positron annihilation era can 
easily be calculated from the temperature of the present cosmic background 
radiation of $T(t_0){\approx}2.73$ K (where the reference epoch $t_0$ has been 
chosen to represent the present era). This calculation yields that the 
quasi-metric universe is (much) older than the big bang universe of the same 
temperature. 

On the other hand, for very early epochs, when temperatures were high enough to 
include a sufficient number of heavy leptons/antileptons or hadrons/antihadrons
in thermal equilibrium, the contribution to $\beta(T)$ from photon energy
density can be neglected. For this case, equation (32) yields ${\beta}(T)
{\lesssim}0.75$. For later epochs, before the era of electron-positron 
annihilation has started but for temperatures low enough so that the abundance 
of muons is negligible (a few MeV), the plasma consisted of photons, electrons,
positrons and neutrinos/antineutrinos (3 types) in thermal equilibrium. The 
temperature at this epoch was sufficiently high (i.e., 
$k_{\rm B}T{\gg}m_{{\rm f}_i}c^2$) to neglect the contribution from fermion mass
$m_{{\rm f}_i}$ to fermion energy density ${\varepsilon}_{{\rm f}_i}$, yielding 
${\varepsilon}_{{\rm f}_i}{\propto}T^4$ (assuming $k_{\rm B}
T{\gg}{\mu}_{{\rm f}_i}$). Then ${\varepsilon}_{{\rm f}_i} \approx 
{\frac{7}{16}}g_i{\varepsilon}_{\gamma}$ follows from equation (28) (by
evaluating the integral neglecting mass and chemical potential). This means 
that the contribution from electron/positron mass to $\beta(T)$ can be 
neglected (as can the contributions from $I_{e^-}$ and $I_{e^+}$), so that to a 
good approximation, ${\beta}(T) \approx {\frac{121}{156}} \approx 0.78$
for this epoch. For later epochs and lower temperatures contributions from
electron/positron mass (and from $I_{e^-}$, $I_{e^+}$) to $\beta(T)$ cannot be 
neglected. However, as long as the neutrinos were still in thermal equilibrium,
$\beta(T)$ cannot increase above a maximum value of ${\frac{95}{116}}
{\approx}0.82$ found by neglecting contributions to $\beta(T)$ from 
electron/positron energy density altogether.

For even lower temperatures, neutrino decoupling starts and leads to major 
changes in $\beta(T)$. As explained in section 3.4; in quasi-metric cosmology
neutrino decoupling proceeds in two stages, where the second stage has no 
counterpart in standard cosmology. At the first stage, some neutrinos fall out 
of thermodynamical equilibrium even though most are in good thermal contact 
with the photon plasma. This means that $\beta(T)$ will increase. As 
temperatures drop even further, the second stage of neutrino decoupling will 
proceed and the production of new neutrinos will effectively cease. This means 
that $\beta(T)$ will first decrease but later it will rise again as even more
neutrinos lose thermal contact with the plasma. Finally $\beta(T){\approx}1$ 
when all electron-positron pairs have annihilated.

In standard cosmology, one may estimate whether or not a particle species is in
good thermal contact with the photons by comparing particle interaction rates
to the expansion rate. More precisely, the particle species actually decouples 
thermally from the photons when the relevant interaction rates become smaller
than the relative change of plasma temperature ${\mid}{\dot T}{\mid}/T$. 
Besides, if heating due to net particle-antiparticle annihilation rates can be
neglected, for a relativistic plasma we have that relative change of plasma
temperature ${\mid}{\dot T}{\mid}/T{\approx}H$ in standard cosmology. Thus to 
estimate the epoch and temperature of thermal decoupling, it is sufficient to 
consider the quantity ${\Gamma}/H$, where ${\Gamma}=n{\langle}{\sigma}{\mid}
v{\mid}{\rangle}$ is the interaction rate per particle [5]. (Here $n$ is the 
number density of target particles, ${\sigma}$ is the interaction cross 
section, ${\mid}v{\mid}$ is the relative speed of the reacting particles and 
${\langle}{\sigma}{\mid}v{\mid}{\rangle}$ is the thermal average of said 
quantities.) For neutrinos, such an estimate [4] yields that ${\Gamma}_{\rm weak}
/H{\approx}{\Big (}{\frac{k_{\rm B}T}{{\rm 1.5{\ }MeV}}}{\Big )}^3$, so for 
temperatures higher than about ${\rm 1.5{\ }MeV}/k_{\rm B}$, neutrinos are in 
good thermal contact with the plasma. However, when the temperature drops below
this value, the neutrinos start to decouple from the plasma and soon lose 
thermal contact with it (this happens before the era of electron-positron 
annihilation, so the assumption ${\mid}{\dot T}{\mid}/T=H$ will hold to a good 
approximation). The neutrinos will then remain thermal but with a temperature 
that diverges from that of the plasma during the epoch of electron-positron 
annihilation. 

In quasi-metric cosmology the relative temperature change of the cosmic plasma 
is found from equations (32) and (48). For the relevant epochs of neutrino 
decoupling in quasi-metric cosmology, this means that we should in principle 
use (a somewhat modified version of) the criterion 
${\Gamma}T/{\mid}{\dot T}{\mid}{\sim}1$ to estimate the temperature of and the 
epoch when the neutrinos finally decouple from the photon plasma in 
quasi-metric space-time. However, since we always have
${\beta}(T){\lesssim}1$, for an estimate it is sufficient to use the criterion
${\Gamma}/H{\sim}1$. But before we do any calculations, we should notice two 
things. First, since the quasi-metric universe is much older than the standard 
big bang universe for a fixed temperature in the relvant range, the 
quasi-metric value of ${\mid}{\dot T}{\mid}/T<H=1/t$ is much smaller than for 
the standard big-bang universe and so is $\Gamma$ for the epoch when 
${\Gamma}/H{\sim}1$. That is, for the quasi-metric universe, neutrino 
decoupling happens at a later epoch and at a lower temperature than for the 
standard big bang universe. Second, due to low temperatures and slow reaction 
rates, neutrino decoupling is a much more gradual process than in standard 
cosmology, so it may be expected that the resulting decoupled neutrino phase 
space distribution will be {\em non-thermal}. Moreover, once the neutrinos have 
completely decoupled, their phase space distribution will be maintained with 
constant neutrino energies. The reason for this that neutrinos are material 
particles (i.e., as long as no neutrino mass eigenstate is null), so once 
decoupled, their thermodynamical properties will evolve in accordance with 
equation (22). Thus quasi-metric theory predicts the existence of a non-thermal
neutrino background with an average neutrino energy much higher than that 
corresponding to the temperature (${\sim}{\rm 1.96{\ }K}$ [5]) 
of its counterpart in standard big bang theory. But as we shall see later, this
predicted relic neutrino background is ruled out from solar neutrino 
experiments. Said prediction is not absolute though, since it is assumed that 
no neutrino mass eigenstate is null and that the massive mass eigenstates do 
not decay over cosmic time periods. That is, if the lightest neutrino mass 
eigenstate is null and the two massive neutrino eigenstates decay into the null
eigenstate, the neutrino energy will be redshifted as if all neutrino mass 
eigenstates were null and the resulting relic neutrino background will then 
have an energy density of the same order as the cosmic microwave background.
Such a low-energy relic neutrino background would be unobservable with today's
experimental techniques.
\subsection{Thermally averaged cross sections}
We start by noting that for the relevant temperature range, the 
electron-positron annihilation process shown in equation (39) is much more 
effective in producing neutrinos than is the photoneutrino process due to a 
smaller cross section for the latter. We will therefore neglect the 
photoneutrino process for the rest of this paper. Consequently, to keep 
neutrinos (almost) in thermodynamical equilibrium, in a comoving volume $V$ 
said annihilation process must be capable of producing at least a minimum 
number rate of neutrinos plus antineutrinos given by $2V{\Gamma}_{\rm ann}
^{\rm net}(t{\leq}t_{\rm C}) \sum_i n^{\rm eff}_{{\nu}_i}{\approx}6Vn^{\rm eff}_{{\nu}}
{\Gamma}_{\rm ann}^{\rm net}(t{\leq}t_{\rm C})$, where ${\Gamma}_{\rm ann}^{\rm net}
(t{\leq}t_{\rm C})$ is obtained from equation (45) by setting ${\dot h}=0$. But 
the annihilation process can at best produce a net number rate of 
neutrino-antineutrino pairs given by 
$Vn_e^2 \sum_i {\langle}{\sigma}_{{\rm ann}_i}{\mid}v{\mid}{\rangle}$, where 
${\sigma}_{{\rm ann}_i}$ is the annihilation cross-section for production of 
neutrino-antineutrino pairs of type $i$. This means that, when eventually 
$3n^{\rm eff}_{{\nu}}{\Gamma}_{\rm ann}^{\rm net}(t{\leq}t_{\rm C}){\sim}n^2_e \sum_i 
{\langle}{\sigma}_{{\rm ann}_i}{\mid}v{\mid}{\rangle}$, the second stage of 
neutrino decoupling will start although there may still be good thermal contact
between the neutrinos and the cosmic plasma via the scattering reactions shown 
in equation (40).

Furthermore, since the relevant temperature range corresponds to energies much 
smaller than the masses of the intermediate bosons $W^{\pm}$ and $Z^0$ 
describing annihilation and scattering reactions in electroweak theory, we can 
use Fermi theory to estimate the corresponding cross sections. Since the 
$W^{\pm}$ bosons are relevant in the weak interaction processes involving 
${\nu}_e$ and ${\bar {\nu}}_e$ only, such processes have different cross 
sections than those involving ${\nu}_{\mu}$, ${\bar {\nu}}_{\mu}$, ${\nu}_{\tau}$ 
and ${\bar {\nu}}_{\tau}$. On the other hand, the cross sections involving 
${\nu}_{\mu}$ are equal to those involving ${\nu}_{\tau}$ (and similarly for
cross sections involving ${\bar {\nu}}_{\mu}$ and ${\bar {\nu}}_{\tau}$). 
Moreover, we have that ${\sigma}_{e^+{\bar {\nu}}_e{\rightarrow}e^+{\bar {\nu}}_e}=
{\sigma}_{e^-{\nu}_e{\rightarrow}e^-{\nu}_e}$ etc. Also, since the neutrinos are
extremely relativistic, we have that ${\langle}{\sigma}^{\rm eff}_{\rm scat}{\mid}v
{\mid}{\rangle}={\langle}{\sigma}^{\rm eff}_{\rm scat}{\rangle}c$ for the 
scattering reactions. This means that ${\Gamma}^{\rm eff}_{\rm scat}$ can be 
defined by the expression (defining the average scattering rate per particle)
\eqa
{\Gamma}^{\rm eff}_{\rm scat}{\equiv}{\frac{1}{3}}
n_ec[{\langle}{\sigma}^{\rm eff}_{e^-{\nu}_e
{\rightarrow}e^-{\nu}_e}{\rangle}+2{\langle}{\sigma}^{\rm eff}_{e^-{\nu}_{\mu}
{\rightarrow}e^-{\nu}_{\mu}}{\rangle}+{\langle}
{\sigma}^{\rm eff}_{e^-{\bar {\nu}}_e{\rightarrow}e^-{\bar {\nu}}_e}{\rangle}+
2{\langle}{\sigma}^{\rm eff}_{e^-{\bar {\nu}}_{\mu}{\rightarrow}e^-
{\bar {\nu}}_{\mu}}{\rangle}].
\ena
Moreover, from the discussion at the beginning of this section we find the
criterion
\eqa
G_{\rm ann}{\equiv}{\frac{n^2_e \sum_i 
{\langle}{\sigma}_{{\rm ann}_i}{\mid}v{\mid}{\rangle}}
{3n^{\rm eff}_{\nu}{\Gamma}_{\rm ann}^{\rm net}(t_{\rm C})}}
={\frac{n_e^2t_{\rm C}[{\langle}{\sigma}_{e^-e^+{\rightarrow}
{\nu}_e{\bar {\nu}}_e}{\mid}v{\mid}{\rangle}
+2{\langle}{\sigma}_{e^-e^+{\rightarrow}{\nu}_{\mu}{\bar {\nu}}_{\mu}}
{\mid}v{\mid}{\rangle}]}{3n_{\nu}^{\rm eff}{\Big [}3(1-{\beta}(T_{\rm C}))+
{\frac{2t_{\rm C}}{3{\zeta}(3)}} 
 \int_0^{{\frac{E_{\rm D}(t_{\rm C})}{k_{\rm B}T_{\rm C}}}}{\frac{x^2dx}
{1+e^x}}{\Gamma}^{\rm eff}_{\rm scat}{\Big ]}}}{\approx}1,
\ena
for estimating (given the temperature $T_{\rm C}$) the neutrino decoupling 
threshold energy $E_{\rm D}(t_{\rm C})$ and the epoch $t_ {\rm C}$ where the second
stage of neutrino decoupling begins. (A second equation relating these 
quantities can be found by applying equation (51) below.)

However, in order to set up a general criterion for when neutrinos with a 
given energy $E_{\rm D}$ decouple from the photon plasma, it is necessary to use
nonaveraged cross sections rather than the thermally averaged ones used in 
equation (50). So, just by using a rate ${\Gamma}_{\rm scat}$ as the nonaveraged 
counterpart to ${\Gamma}^{\rm eff}_{\rm scat}$, we are able to set up the criterion
\eqa
G_{\rm scat}{\equiv}{\frac{{\Gamma}_{\rm scat}T}
{{\mid}{\dot{T}}{\mid}}}={\frac{n_ect}{3{\beta}(T)}}{\Big [}
{\sigma}^{\rm eff}_{e^-{\nu}_e{\rightarrow}e^-{\nu}_e}+
2{\sigma}^{\rm eff}_{e^-{\nu}_{\mu}{\rightarrow}e^-{\nu}_{\mu}}
+{\sigma}^{\rm eff}_{e^-{\bar {\nu}}_e{\rightarrow}e^-{\bar {\nu}}_e}
+2{\sigma}^{\rm eff}_{e^-{\bar {\nu}}_{\mu}{\rightarrow}e^-
{\bar {\nu}}_{\mu}}{\Big ]}{\approx}1,
\ena
for estimating the energy $E_{\rm D}$ (at given temperature $T$ and epoch $t$) 
where the weak interactions (involving all the relevant neutrino scattering 
reactions) become ineffective to maintain thermal contact between the neutrinos
with energy equal to $E_{\rm D}$ and the cosmic photon plasma. That is, when 
$G_{\rm scat}$ drops below unity, the weak interactions for neutrinos with energy
equal to $E_{\rm D}$ become slow compared to the relative temperature change of 
the plasma. Now the general expression for the weak cross section 
${\sigma}_{\rm scat}$ valid for the scattering reactions relevant for equations 
(49) and (51) is (see, e.g., [6])
\eqa
{\sigma}_{\rm scat}{\equiv}{\frac{G_{\rm F}^2m_eE_{\rm max}}
{2{\pi}{\hbar}^4c^2}}{\Big [}(g_{\rm v}+g_{\rm a})^2+(g_{\rm v}-g_{\rm a})^2
(1-{\frac{E_{\rm max}}{E_{\nu}}}+{\frac{E_{\rm max}^2}{3E_{\nu}^2}})+
(g_{\rm a}^2-g_{\rm v}^2){\frac{m_ec^2E_{\max}}{2E_{\nu}^2}}{\Big ]},
\ena
where $G_{\rm F}$ is the Fermi constant and $E_{\nu}$ is the energy of the 
incoming neutrino. (Numerically, $G_{\rm F}{\equiv}1.1664{\times}10^{-5}{\hbar}^3
c^3{\ }{\rm GeV}^{-2}=1.4361{\times}10^{-49}{\ }{\rm cm}^5{\rm g}/{\rm s}^2$.) 
Furthermore, $g_{\rm v}$ and $g_{\rm a}$ are respectively vectorial and axial 
coupling constants and $E_{\rm max}$ is the maximum recoil kinetic energy of the 
target electron, i.e.,
\eqa
E_{\rm max}{\equiv}{\frac{2E_{\nu}^2}{m_ec^2+2E_{\nu}}}, \qquad
g_{\rm v}=2{\sin}^2{\theta}_{\rm w}{\pm}{\frac{1}{2}}, \qquad 
g_{\rm a}={\pm}{\frac{1}{2}}, \qquad {\sin}^2{\theta}_{\rm w}{\approx}0.23,
\ena
where ${\theta}_{\rm w}$ is the Weinberg angle. Here the plus sign is used in
the expressions for the coupling constants for reactions involving ${\nu}_e$,
while the minus sign is used for reactions involving ${\nu}_{\mu}$ and
${\nu}_{\tau}$. (For reactions involving antineutrinos, just let
$g_{\rm a}{\rightarrow}-g_{\rm a}$.) Equations (52) and (53) straightforwardly
yield
\eqa
{\sigma}^{\rm eff}_{e^-{\nu}_e{\rightarrow}e^-{\nu}_e}+
2{\sigma}^{\rm eff}_{e^-{\nu}_{\mu}{\rightarrow}e^-{\nu}_{\mu}}
+{\sigma}^{\rm eff}_{e^-{\bar {\nu}}_e{\rightarrow}e^-{\bar {\nu}}_e}
+2{\sigma}^{\rm eff}_{e^-{\bar {\nu}}_{\mu}{\rightarrow}e^-
{\bar {\nu}}_{\mu}} \nonumber \\
={\frac{G_{\rm F}^2m_e}{{\pi}c^2{\hbar}^4}}
{\frac{E_{\rm D}^2}{m_ec^2+2E_{\rm D}}}{\Big [}(24{\sin}^4{\theta}_{\rm w}-
4{\sin}^2{\theta}_{\rm w}+3){\Big (}2-{\frac{2E_{\rm D}}{m_ec^2+2E_{\rm D}}}+
{\frac{4E_{\rm D}^2}{3(m_ec^2+2E_{\rm D})^2}}{\Big )}
\nonumber \\
-4{\sin}^2{\theta}_{\rm w}
(6{\sin}^2{\theta}_{\rm w}-1){\frac{m_ec^2}{m_ec^2+2E_{\rm D}}}{\Big ]}.
\ena
Similarly, the expression for the weak cross section ${\sigma}_{\rm ann}$ 
evaluated in the center-of-mass (c.m.) frame and valid for the annihilation 
reactions producing neutrinos shown in equation (39) is given by [7]
\eqa
{\sigma}_{\rm ann}{\equiv}{\frac{G_{\rm F}^2m_e^2c}
{6{\pi}{\hbar}^4{\mid}v{\mid}}}{\Big {\{}}(g_{\rm v}^2+g_{\rm a}^2){\Big [}
{\Big (}{\frac{2E_e}{m_ec^2}}{\Big )}^2-1{\Big ]}+
3(g_{\rm v}^2-g_{\rm a}^2){\Big [}2-{\frac{m_e^2c^4}{2E_e^2}}{\Big ]}
{\Big {\}}},
\ena
where $E_e=E_{\nu}$ is the energy of the incoming electron/positron, or
equivalently, the energy of the produced neutrino/antineutrino in the c.m. 
frame. (Also, ${\mid}v{\mid}$ is the relative speed of the annihilating
electron-positron pair.)

Now the (effective) thermally averaged cross sections
${\langle}{\sigma}^{\rm eff}_{\rm scat}{\rangle}c$ and
${\langle}{\sigma}_{\rm ann}{\mid}v{\mid}{\rangle}$
can be found from equations (52) and (55), respectively. The results are
\eqa
{\langle}{\sigma}^{\rm eff}_{\rm scat}{\rangle}c={\frac{G_{\rm F}^
2m_ek_{\rm B}T}{{\pi}{\hbar}^4cI_0}}{\Big {\{}}2(g_{\rm v}^2+g_{\rm a}^2)
I_1{\Big (}{\frac{m_ec^2}{k_{\rm B}T}}{\Big )} \nonumber \\
+(g_{\rm v}-g_{\rm a})^2{\Big [}{\frac{4}{3}}
I_4{\Big (}{\frac{m_ec^2}{k_{\rm B}T}}{\Big )}
-2I_3{\Big (}{\frac{m_ec^2}{k_{\rm B}T}}{\Big )}{\Big ]}+
(g_{\rm a}^2-g_{\rm v}^2){\frac{m_ec^2}{k_{\rm B}T}}
I_2{\Big (}{\frac{m_ec^2}{k_{\rm B}T}}{\Big )}{\Big {\}}},
\ena
where
\eqa
I_0{\equiv}{\int}_{{\!}{\!}{\!}{\!}{\frac{E_{\rm D}}{k_{\rm B}T}}}^{\infty}
{\frac{x^2dx}{1+e^x}}, \quad
I_1(y){\equiv}{\int}_{{\!}{\!}{\!}{\!}{\frac{E_{\rm D}}{k_{\rm B}T}}}^{\infty}
{\frac{x^4dx}{(y+2x)(1+e^x)}},
\quad
I_2(y){\equiv}{\int}_{{\!}{\!}{\!}{\!}{\frac{E_{\rm D}}{k_{\rm B}T}}}^{\infty}
{\frac{x^4dx}{(y+2x)^2(1+e^x)}}, \nonumber \\
I_3(y){\equiv}{\int}_{{\!}{\!}{\!}{\!}{\frac{E_{\rm D}}{k_{\rm B}T}}}^{\infty}
{\frac{x^5dx}{(y+2x)^2(1+e^x)}}, \qquad
I_4(y){\equiv}{\int}_{{\!}{\!}{\!}{\!}{\frac{E_{\rm D}}{k_{\rm B}T}}}^{\infty}
{\frac{x^6dx}{(y+2x)^3(1+e^x)}},
\ena
and
\eqa
{\langle}{\sigma}_{\rm ann}{\mid}v{\mid}{\rangle}={
\frac{G_{\rm F}^2m_e^2c}
{6{\pi}{\hbar}^4}}{\Big {\{}}2(g_{\rm v}^2+g_{\rm a}^2){\Big [}
{\frac{4m_e^3c^3}{{\pi}^2{\hbar}^3n_e}}
I_5{\Big (}{\frac{m_ec^2}{k_{\rm B}T}}{\Big )}-1{\Big ]} 
\nonumber \\
+3(g_{\rm v}^2-g_{\rm a}^2){\Big [}2-{\frac{m_e^3c^3}{2{\pi}^2{\hbar}^3n_e}}
I_6{\Big (}{\frac{m_ec^2}{k_{\rm B}T}}{\Big )}{\Big ]}{\Big {\}}},
\ena
where
\eqa
I_5(y){\equiv}{\int}_{{\!}{\!}{\!}{\!}1}^{\infty}
{\frac{x^3{\sqrt{x^2-1}}dx}{1+{\exp}(yx)}},
\qquad
I_6(y){\equiv}{\int}_{{\!}{\!}{\!}{\!}1}^{\infty}
{\frac{{\sqrt{x^2-1}}dx}{x[1+{\exp}(yx)]}}.
\ena
Note that the lower integration limit for the integrals in equation (57) is
non-zero since neutrinos with energy below $E_{\rm D}$ are decoupled from
the cosmic plasma.

Using equations (53) and (56) we then find
\eqa
{\langle}{\sigma}^{\rm eff}_{e^-{\nu}_e{\rightarrow}e^-{\nu}_e}{\rangle}+
2{\langle}{\sigma}^{\rm eff}_{e^-{\nu}_{\mu}{\rightarrow}e^-{\nu}_{\mu}}
{\rangle}+{\langle}{\sigma}^{\rm eff}_{e^-{\bar {\nu}}_e{\rightarrow}e^-
{\bar {\nu}}_e}{\rangle}+2{\langle}{\sigma}^{\rm eff}_{e^-{\bar {\nu}}_{\mu}
{\rightarrow}e^-{\bar {\nu}}_{\mu}}{\rangle}
 \nonumber \\ 
={\frac{2G_{\rm F}^2m_ek_{\rm B}T}{{\pi}{\hbar}^4c^2I_0}}
{\Big {\{}}(24{\sin}^4{\theta}_{\rm w}-4{\sin}^2{\theta}_{\rm w}+3){\Big [}
I_1{\Big (}{\frac{m_ec^2}{k_{\rm B}T}}{\Big )}-
I_3{\Big (}{\frac{m_ec^2}{k_{\rm B}T}}{\Big )}+{\frac{2}{3}}
I_4{\Big (}{\frac{m_ec^2}{k_{\rm B}T}}{\Big )}{\Big ]} \nonumber \\
+2{\sin}^2{\theta}_{\rm w}(1-6{\sin}^2{\theta}_{\rm w})
{\frac{m_ec^2}{k_{\rm B}T}}I_2{\Big (}{\frac{m_ec^2}{k_{\rm B}T}}{\Big )}
{\Big {\}}},
\ena
and equation (58) yields
\eqa
{\langle}{\sigma}_{e^-e^+{\rightarrow}{\nu}_e{\bar {\nu}}_e}
{\mid}v{\mid}{\rangle}+
2{\langle}{\sigma}_{e^-e^+{\rightarrow}{\nu}_{\mu}{\bar {\nu}}_{\mu}}
{\mid}v{\mid}{\rangle} \nonumber \\ 
={\frac{G_{\rm F}^2m_e^2c}{6{\pi}{\hbar}^4}}
{\Big {\{}}(24{\sin}^4{\theta}_{\rm w}-4{\sin}^2{\theta}_{\rm w}+3){\Big [}
{\frac{4m_e^3c^3}{{\pi}^2{\hbar}^3n_e}}
I_5{\Big (}{\frac{m_ec^2}{k_{\rm B}T}}{\Big )}-1{\Big ]} \nonumber \\
+6{\sin}^2{\theta}_{\rm w}(6{\sin}^2{\theta}_{\rm w}-1){\Big [}2-
{\frac{m_e^3c^3}{2{\pi}^2{\hbar}^3n_e}}
I_6{\Big (}{\frac{m_ec^2}{k_{\rm B}T}}{\Big )}{\Big ]}{\Big {\}}}.
\ena
\subsection{The decoupling temperature $T_{\rm dec}$}
The basic assumption underlying the model approximately describing the 
process of neutrino decoupling in quasi-metric cosmology is given in equation
(42). The quantities $T(t)$, $h(t)$ and $E_{\rm D}(t)$ entering this equation 
are estimated rather than defined from exact formulae. In particular, the
temperature $T_{\rm C}$ does not have an exact value but rather has an 
uncertainity associated with it due to the approximate way it is calculated.
Similarly, one may define a ``decoupling temperature'' $T_{\rm dec}$ as the
temperature where the contribution to ${\beta}(T)$ from the neutrinos becomes
negligible. We thus define this temperature (somewhat arbitrarily) as that at
epoch $t_{\rm dec}$ where said contribution from the neutrinos drops below
${\sim}0.001$ (see fig.~\ref{fig:nbeta} below). The arbitrariness of this 
definition means that $T_{\rm dec}$ cannot have an exact value, however since 
said contribution drops quickly with increasing $t$ for $t{\approx}t_{\rm dec}$,
the uncertainity in the definition does not matter much when estimating 
$t_{\rm dec}$ and $T_{\rm dec}$. Therefore it is still meaningful to speak of the 
decoupling temperature $T_{\rm dec}$ and the decoupling epoch $t_{\rm dec}$. 

Next, we have assumed that the whole process of neutrino decoupling happens in 
a temperature range where $k_{\rm B}T{\ll}m_ec^2$ holds, so that the number 
density of positrons should be much smaller than the number density of photons.
On the other hand, just before $t_{\rm dec}$ the number density of 
heat-conducting electrons/positrons must be much larger than $n_{e^-}-n_{e^+}$, 
i.e., ${\frac{n_{e^-}-n_{e^+}}{n_{\gamma}}}{\ll}
{\frac{n_{e^-}}{n_{\gamma}}}{\ll}1$. That is, we assume that $T_{\rm dec}$ is 
sufficiently high compared to the estimated temperature ${\sim}20$ 
keV$/k_{\rm B}$ where the number density of positrons $n_{e^+}$ becomes 
negligible. This means that to a good approximation, we can set 
$n_{e^+}{\approx}n_{e^-}$ so that we can neglect the associated chemical 
potential, i.e., we can set ${\mu}_e{\approx}0$ in the relevant temperature 
range. Thus we can neglect the contribution to ${\beta}(T)$ from ${\mu}_e$,
as we have done in the previous section.

To find the unknown functions $T(t)$, $h(t)$ and $E_{\rm D}(t)$ in the relevant
temperature range $T_{\rm C}{\geq}T{\geq}T_{\rm dec}$ we must solve numerically
the coupled set of equations (32) (with ${\beta}(T)$ given from equation (48))
and (45) using the criterion (51) to close the set. Now equations
(32) and (45) are integrodifferential equations rather then ordinary 
differential equations (ODEs), so solving them numerically using MAPLE is not 
straightforward. However, it is possible to approximate the relevant integrals
with other integrals recognized as special functions in MAPLE, approximating
equations (32) and (45) with two coupled ODEs that can be straightforwarly 
solved numerically using MAPLE. Note that $T_{\rm C}$ plays the role of a chosen
boundary value and that for any given choice, $t_{\rm C}$ and $E_{\rm D}(t_{\rm C})$
can be found from the criteria (50) and (51). Moreover, $T_{\rm C}$ must be 
chosen such that the calculated values $T_{\rm dec}$ and $t_{\rm dec}$ agree with 
these values when solving equation (32) with the boundary value $T_0$ at age 
$t_0$ for the present temperature of the cosmic microwave background. See 
appendix A for detailed formulae and further instructions of how to solve said 
ODEs.

\begin{figure}[t]
\begin{center}
\includegraphics*[height=4.8699cm,width=5.0cm,angle=0]{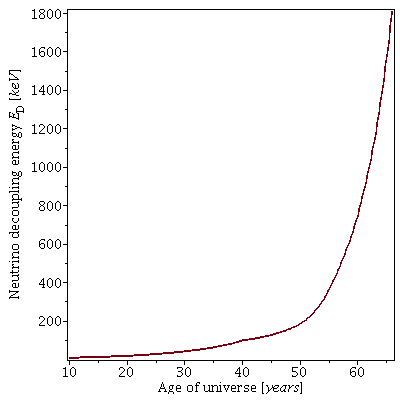}
\caption{\label{fig:deckev} The energy threshold $E_{\rm D}$ of decoupling 
neutrinos versus age.}
\end{center}
\end{figure}

\begin{figure}[t]
\begin{center}
\includegraphics*[height=4.8699cm,width=5.0cm,angle=0]{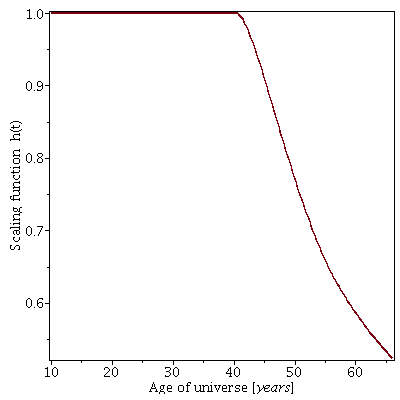}
\caption{\label{fig:nscale} The scaling function h(t) for neutrino densities.}
\end{center}
\end{figure}

\begin{figure}[t]
\begin{center}
\includegraphics*[height=4.8699cm,width=5.0cm,angle=0]{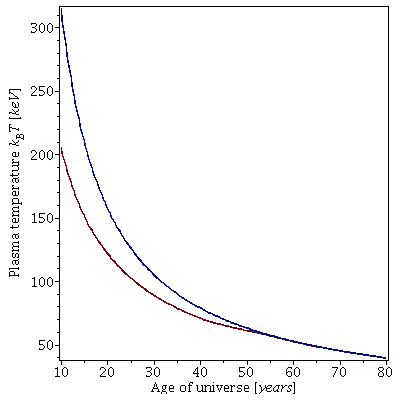}
\caption{\label{fig:tempkev} Plasma temperature versus age. Plasma 
temperature versus age for a pure photon plasma is included for comparison 
(top curve).}
\end{center}
\end{figure}

The results of the numerical procedure are $T_{\rm C}{\approx}70$ keV$/k_{\rm B}$
and $t_{\rm C}{\approx}40$ yr yielding $T_{\rm dec}{\approx}53$ keV$/k_{\rm B}$ 
corresponding to an age of $t_{\rm dec}{\approx}59$ yr. This means that there 
are about 19 years separating the epochs $t_{\rm C}$ and $t_{\rm dec}$. We note 
that $h(t_{\rm dec}){\approx}0.59$, $E_{\rm D}(t_{\rm C}){\approx}98$ keV and 
$E_{\rm D}(t_{\rm dec}){\approx}673$ keV, so the value of $E_{\rm D}$  increases 
with a factor about 7 during said time interval. Moreover, the estimate for 
$T_{\rm dec}$ is not very sensitive to the value of the input cross sections, 
since multiplying all scattering cross sections with a factor of 2 yields only 
a change of about $5{\%}$ in the estimates for $T_{\rm C}$, $T_{\rm dec}$, 
$t_{\rm C}$, $t_{\rm dec}$, $h(t_{\rm dec})$ and $E_{\rm D}(t_{\rm dec})$, while the 
value of $E_{\rm D}(t_{\rm C})$ changes with about $20{\%}$. Thus a less tight 
estimate of $T_{\rm dec}{\sim}50-60$ keV$/k_{\rm B}$ for the decoupling 
temperature seems reasonable. Some of these results are illustrated 
in figs.~\ref{fig:deckev},~\ref{fig:nscale},~\ref{fig:tempkev} and
~\ref{fig:nbeta}, respectively. However, a note of warning is required 
concerning the validity of said results. Since 
$E_{\rm D}(t_{\rm C}){\sim}k_{\rm B}T_{\rm C}$, this could indicate that the model
{\em ansatz} given by equation (42) breaks down already for $t<t_{\rm C}$,
so that a more realistic decoupling temperature could be $T_{\rm dec}>T_{\rm C}$.
If so, the function ${\beta}(t)$ should not drop suddenly just after
$t=t_{\rm C}$ as shown in fig. 4, but rather be strictly increasing and lying
closer to its counterpart where the contribution from neutrinos has been 
omitted.
\begin{figure}[t]
\begin{center}
\includegraphics*[height=4.8699cm,width=5.0cm,angle=0]{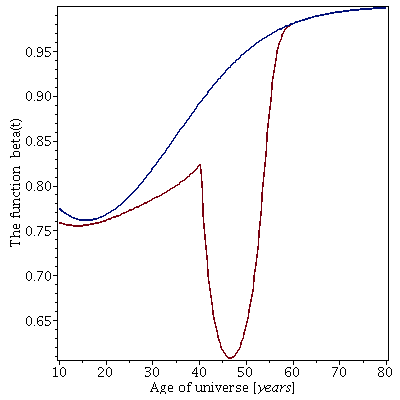}
\caption{\label{fig:nbeta} The function ${\beta}(t)$. The function 
${\beta}(t)$ without the contribution from neutrinos is included for 
comparison (top curve).}
\end{center}
\end{figure}
Also, in this case the neutrino distributions should be significantly 
depleted for energies above about $3 k_{\rm B}T_{\rm dec}$.

We could also try to estimate the temperature and age at neutrino decoupling 
for a ``coasting power-law'' cosmology (CPLC), i.e., a Robertson-Walker 
model where the Friedmann equations are made irrelevant by postulating a scale 
factor $a(t){\propto}t^{\alpha}$ with ${\alpha}=1$. See, e.g., [8] for more 
details on such a cosmology. For a general power-law cosmology with arbitrary
$\alpha$, standard thermodynamics applies, so entropy conservation for the 
total entropy yields $s(t)t^{3{\alpha}}={\rm constant}$. Assuming that the 
neutrinos share most of the heating coming from net electron-positron 
annihilation so that the neutrinos and the photon plasma have approximately 
the same temperature, equation (37) yields (assuming massless neutrinos and
${\mu}_{{\nu}_i}{\equiv}0$)
\eqa
T{\approx}{\frac{t_0^{\alpha}T_0}{t^{\alpha}}}
{\Big (}{\frac{{\frac{4}{3}}{\varepsilon}_{\gamma}+
\sum_i{\Big [}2{\varepsilon}_{{\nu}_i}^{\rm dec}(t_0)
+2p_{{\nu}_i}^{\rm dec}(t_0){\Big ]}{\frac{T^4}{T_0^4}}}
{{\frac{4}{3}}{\varepsilon}_{\gamma}+
\sum_i{\Big [}{\varepsilon}_{{\rm p}_i}
+{\varepsilon}_{{\rm {\bar p}}_i}+p_{{\rm p}_i}+p_{{\rm {\bar p}}_i}-
{\mu}_i(n_{{\rm p}_i}-n_{{\rm {\bar p}}_i})+2{\varepsilon}_{{\nu}_i}
+2p_{{\nu}_i}{\Big ]}}}{\Big )}^{\frac{1}{3}},
\ena
where $2 \sum_i{\varepsilon}_{{\nu}_i}^{\rm dec}(t_0)$ is the energy density of the
relic neutrino/antineutrino background at the present era. However, since the 
approach used in this paper is inconsistent with standard thermodynamics, 
equation (62) (with ${\alpha}=1)$ is not consistent with equation (32). This 
means that it is not meaningful to apply the model given by equation (42) to 
CPLCs. To estimate the decoupling temperature $T_{\rm dec}$ and decoupling era 
$t_{\rm dec}$ for a CPLC, we may rather try the usual approach assuming 
instantaneous decoupling. The decoupling criterion 
${\Gamma}^{\rm eff}_{\rm scat}T/{\mid}{\dot T}{\mid}{\sim}1$ then yields (using 
equations (49) and (62)) $T_{\rm dec}{\approx}64$ keV$/k_{\rm B}$ and 
$t_{\rm dec}{\approx}52$ years. This result is not very different from that 
obtained for the quasi-metric model. Finally, we note that the decoupling 
temperature for a coasting cosmology was estimated to be about $75$ 
keV$/k_{\rm B}$ (using $H(t_0)=65{\ }{\frac{{\rm km}}{{\rm sMpc}}}$) in [8], 
using a criterion ${\Gamma}_{\rm weak}/H{\sim}1$ without thermal averaging and 
assuming that this quantity approximately goes as ${\Big[}{\frac{T}{
1.6{\times}10^8{\ }{\rm K}}}{\Big ]}^4{\exp}(-m_e c^2/k_{\rm B}T)$ for the 
relevant temperature range.
\subsection{The neutrino distributions after decoupling}
In the previous section we found that the process of neutrino decoupling is
estimated to take several years. It is expected that such a gradual neutrino 
decoupling will result in nonthermal neutrino number and energy densities of 
the decoupled neutrinos. Since the cosmic expansion does not redshift momenta 
of free neutrinos (if no mass eigenstate is null), the neutrino number 
and energy density distributions just after decoupling directly yield the 
predicted relic neutrino distributions today since the decoupled densities 
evolve as $t^{-3}$ once all neutrinos have decoupled.

\begin{figure}[t]
\begin{center}
\includegraphics*[height=4.8699cm,width=5.0cm,angle=0]{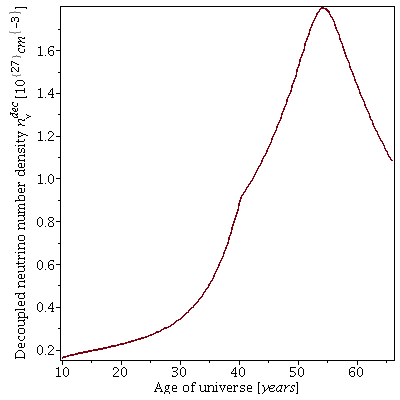}
\caption{\label{fig:nden} Decoupled neutrino number density $n^{\rm dec}_{\nu}$
versus age.}
\end{center}
\end{figure}

To begin with it is straightforward to estimate the number density 
$n_{\nu}^{\rm dec}(t)$ as a function of time by integrating equation (44) 
numerically. (See appendix A for details how to do this.) The result is plotted
in fig.~\ref{fig:nden}. In particular we find $n_{\nu}^{\rm dec}(t_{\rm dec})
{\approx}1.5{\times}10^{27}$ cm$^{-3}$. One may also estimate the number density
of neutrinos decopled via interactions only by omitting the last term of 
equation (44). The result found is that at epoch $t_{\rm C}$, about 44${\%}$ of 
all neutrinos have decoupled via interactions, while at epoch $t_{\rm dec}$,
this number has diminished to about 17${\%}$. Thus the low-energy part of the 
decoupled neutrinos has a significant number of neutrinos which decoupled via 
interaction, while the high-energy part consists almost only of neutrinos that
have ``frozen in''. This indicates that the approximate model given by 
equations (42) and (51) overestimates thermal contact via elastic neutrino 
scattering for the later stages of neutrino decoupling such that the estimated 
value of $T_{\rm dec}$ may be too low. The existence of approximately thermal 
high-energy tails of the decoupled neutrino distributions is also thrown 
further in doubt.

\begin{figure}[t]
\begin{center}
\includegraphics*[height=4.8699cm,width=5.0cm,angle=0]{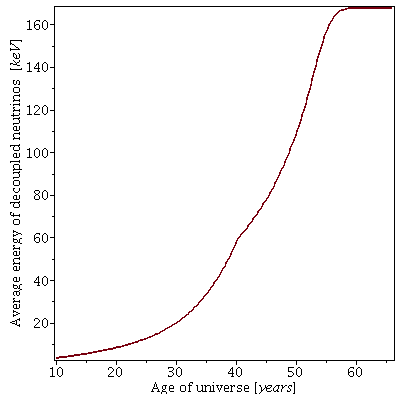}
\caption{\label{fig:avg} Average energy of decoupled neutrinos 
${\langle}E^{\rm dec}_{\nu}{\rangle}$ versus age.}
\end{center}
\end{figure}

Moreover one may estimate the energy density ${\varepsilon}_{\nu}^{\rm dec}(t)$ as 
a function of time by integrating equation (47) numerically. But it is more
interesting to estimate the average neutrino energy per particle for the
decoupled neutrinos, i.e., ${\langle}E^{\rm dec}_{\nu}{\rangle}=
{\frac{{\varepsilon}_{\nu}^{\rm dec}}{n^{\rm dec}_{\nu}}}$. The result is plotted as 
a function of time in fig.~\ref{fig:avg}. After all neutrinos have decoupled,
they have an average energy of about 168 keV. On the other hand there is not 
enough information to calculate the number density distribution 
$n^{\rm dec}_{\nu}(E_{\nu}^{\rm dec})$ or the energy density distribution
${\varepsilon}^{\rm dec}_{\nu}(E_{\nu}^{\rm dec})$.

Once the neutrinos have decoupled, they will stream freely outwards from the 
last scattering surface with a speed close to the speed of light. Just after
the last scattering the neutrinos are in their flavour eigenstates.
However, the flavour eigenstates may be described as wave-packets consisting
of a superposition of mass eigenstates with different masses. The different 
mass eigenstates ${\nu}_j$ (with mass $m_j$) will propagate with different 
speeds $v_j$, leading to wave-packet separation and decoherence. This means
that, soon after decoupling, the decoupled neutrinos will propagate as separate
mass eigenstates and not as flavour eigenstates.
\subsection{The predicted relic neutrino background}
The main result of the previous section was the prediction from quasi-metric 
cosmology that the cosmological neutrino background just after decoupling would
be nonthermal. Moreover, if we assume the approximate validity of equation (42)
throughout neutrino decoupling, the negligible decoupling via interaction
towards the end indicates the existence of a ``frozen in'' high-energy tail of 
the neutrino energy distribution. This high-energy tail should be approximately
thermal, so that the non-thermal features apply mainly to the low-energy 
part of the neutrino energy distribution. Furthermore, over time this neutrino 
background will keep its (nonthermal) characteristic with all neutrinos having 
the same energy they had just after decoupling, yielding a ``high-energy'' 
relic neutrino background today. In addition, today's relic neutrino 
background should be in the form of separate neutrino mass eigenstates since 
the heavier mass eigenstates should trail far behind the lightest mass 
eigenstate.

Standard cosmology predicts a cosmic neutrino number density of each flavour 
(neutrinos plus antineutrinos) of about $112$ cm$^{-3}$ and a neutrino relic 
background very close to thermal with an ``effective'' temperature [9] of about 
$1.96$ K. 
Moreover, data from neutrino oscillation experiments yield information on the 
quantities ${\Delta}m_{ij}^2{\equiv}{\mid}m_i^2-m_j^2{\mid}$. That is, 
${\Delta}m_{21}^2c^4=7.50^{+0.18}_{-0.19}{\times}10^{-5}$ eV$^2$ and 
${\Delta}m_{31}^2c^4=2.47^{+0.170}_{-0.067}{\times}10^{-3}$ eV$^2$ (see, e.g. [10]). 
This means, assuming a ``normal'' mass hierarchy, that $m_3c^2>m_2c^2
{\gg}k_{\rm B}T_{\nu}{\sim}1.7{\times}10^{-4}$ eV, i.e., that at least the two 
most massive relic neutrino mass eigenstates will be nonrelativistic. Thus the 
cosmic neutrino background as predicted from standard cosmology should be 
nonrelativistic at the present epoch. On the other hand, quasi-metric cosmology
predicts a relic neutrino backround with properties very different from its 
counterpart in standard cosmology, so a crucial question is if the existence of
this nonstandard background would be consistent with current experimental 
results. To answer that, we must first find the flux of background neutrinos at
the present epoch. Since after neutrino decoupling, ${\beta}(T){\approx}1$ to 
a good approximation, we find from the previous section that the present number 
density of each neutrino mass eigenstate ${\nu}_j$, i.e., $n_{{\nu}_j}(t_0)$, is 
approximately equal to an effective number density of each neutrino flavour 
eigenstate ${\nu}_{\alpha}$ of $n_{{\nu}_{\alpha}}(t_0)={\frac{t_{\rm dec}^3}{t_0^3}}
n_{{\nu}_{\alpha}}(t_{\rm dec}){\approx}129$ cm$^{-3}$. That is, the effective number
density of neutrinos plus antineutrinos of each flavour is about $258$ 
cm$^{-3}$. This is more than twice the number density predicted from standard 
cosmology.

The above neutrino number density predictions from quasi-metric cosmology imply
that the effective flux of relic electron-neutrinos at the present era is 
predicted to be about $cn_{{\nu}_e}(t_0){\approx}3.9{\times}10^{12}$ 
cm$^{-2}$s$^{-1}$. This may be compared to the flux of low-energy 
electron-neutrinos produced in nuclear reactions taking place in the Sun's 
core (as predicted from standard solar models) and measured at the Earth. That 
is, while the total flux of solar neutrinos is predicted to be about 
$6.5{\times}10^{10}$ cm$^{-2}$s$^{-1}$, the flux of low-energy neutrinos is 
predicted to be about $6.0{\times}10^{10}$ cm$^{-2}$s$^{-1}$ and to arise from 
the dominant proton-proton chain. These predictions of solar neutrino fluxes
agree very well with measurements made in the Borexino experiment [11].
On the other hand, the effective estimated total flux of cosmic 
electron-neutrinos coming from quasi-metric cosmology exceeds the estimated 
total flux of solar neutrinos by a factor of about 60, indicating a violent 
conflict with this experiment. Said prediction is also inconsistent with 
experiments (GALLEX/GNO, SAGE) detecting low-energy neutrinos from the Sun
using the reaction $^{71}{\rm Ga}+{\nu}_e{\ }{\rightarrow}{\ }^{71}{\rm Ge}+e^-$.
(In this reaction, the neutrino is required to have a minimum energy of 
${\rm E}_{\nu}^{\rm min}=233$ keV.) In particular, the experiment SAGE measured an
electron-neutrino capture rate consistent with an electron-neutrino flux at the
location of the Earth of about $3.4{\times}10^{10}$ cm$^{-2}$s$^{-1}$ coming from
the proton-proton chain [12]. (The discrepancy with said theoretical result of 
about $6.0{\times}10^{10}$ cm$^{-2}$s$^{-1}$ is explained as an effect due to 
neutrino oscillations.) On the other hand, we may estimate an ``effective'' 
cosmic electron-neutrino flux assuming that relic neutrinos with energy above 
said minimum energy are given approximately by equation (42) with 
$E_{\rm D}{\approx}E^{\rm min}_{\nu}$ corresponding to an age
$t_{\rm min}{\approx}52$ yr and a plasma temperature $T_{\rm min}{\approx}59$ 
keV/$k_{\rm B}$, i.e.,
\eqa
{\Phi}_{{\nu}_e}^{\rm cosm}(E_{\nu}{\geq}{E_{\nu}^{\rm min}}){\approx}
{\frac{t_{\rm min}^3}{t_0^3}}{\frac{h(t_{\rm min})k_{\rm B}^3T_{\rm min}^3}
{2{\pi}^2{\hbar}^3c^2}}{\int}_{{\!}{\!}{\!}{\!}{\frac{E_{\nu}^{\rm min}}
{k_{\rm B}T_{\rm min}}}}^{\infty}{\frac{x^2dx}{1+e^x}}{\approx}
8.7{\times}10^{11}{\ }{\rm cm}^{-2}{\rm s}^{-1},
\ena
demonstrating that the effective flux of cosmic electron neutrinos in the 
relevant energy range is about 13 times the total flux of solar neutrinos. 
This means that the possible existence of the predicted relic cosmic neutrino 
background is strongly inconsistent with experimental data. This conclusion is 
confirmed by calculating the corresponding capture rate $R^{\rm cosm}_{{\nu}_e}$ 
of cosmological electron-neutrinos in a $^{71}{\rm Ga}$-detector. This is given
by (using a similar definition as given in [12] for
$240{\ }{\rm keV}{\leq}E_{\nu}{\leq}733{\ }{\rm keV}$)
\eqa
&R_{{\nu}_e}^{\rm cosm}{\equiv}{\int}_{{\!}{\!}{240{\ }{\rm keV}}}
^{733{\ }{\rm keV}}{\sigma}_{\rm Ga}(E_{\nu})
{\Upsilon}^{\rm cosm}_{{\nu}_e}(E_{\nu})dE_{\nu}, \quad
{\Upsilon}^{\rm cosm}_{{\nu}_e}(E_{\nu}){\equiv}
{\frac{t_{\rm min}^3}{t_0^3}}{\frac{h(t_{\rm min})}{2{\pi}^2{\hbar}^3c^2}}
{\frac{E_{\nu}^2}{[1+{\exp}(E_{\nu}/k_{\rm B}T_{\rm min})]}}, \nonumber \\
&{\sigma}_{\rm Ga}(E_{\nu}){\approx}[13.10+91.29(
{\frac{E_{\nu}}{1 {\ }{\rm MeV}}}-0.24{\ })^{1.157}]{\times}10^{-46} {\ }
{\rm cm}^2, 
\ena
where ${\sigma}_{\rm Ga}(E_{\nu})$ is the estimated cross section for neutrino
capture by $^{71}{\rm Ga}$ given in [12] (we have used the fact that
contributions to $R_{{\nu}_e}^{\rm cosm}$ are negligibile for $E_{\nu}>733$ keV). 
Moreover, ${\Upsilon}^{\rm cosm}(E_{\nu})$ is the estimated differential flux of 
cosmic neutrinos with energy $E_{\nu}$. By inserting the expression for 
${\sigma}_{\rm Ga}(E_{\nu})$ into equation (64) we find a rate of 
$R^{\rm cosm}_{{\nu}_e}{\sim}15{\times}10^{-34}$ s$^{-1}{\equiv}1.5{\times}10^3$ SNU. 
However, the weighted combination of all observational Ga-experiments 
yields a result of only about $66$ SNU [12], to be compared to the calculated 
contribution from solar neutrinos given by 128 SNU (without taking into account
neutrino oscillations). In other words, the existence of a cosmic neutrino 
backround with an approximately thermal high-energy tail with temperature of 
about $50-60$ keV/$k_{\rm B}$ is in violent conflict with gallium experiments
as well. 

However, as mentioned above, one might argue that any thermal high-energy tail 
is expected to be significantly depleted due to severe deviations from
equilibrium towards the end of neutrino decoupling. This could make it possible
in principle (but not very likely), to reduce the density of relic neutrinos 
with energy above 233 keV with the necessary factor of about 1000 or more. On 
the other hand, in the Borexino experiment one has deteced solar neutrinos 
with energies down to 165 keV without finding any excess other than the 
expected effects coming from radiative decay of $^{14}$C naturally occuring in 
the organic detector fluid [11]. For such low energies excess detections due 
to relic neutrinos are expected to be seen anyway, meaning that said conflict 
can be resolved only by invoking non-standard neutrino physics. In particular, 
if neutrinos are {\em unstable}, decaying into massless or ``invisible'' 
particles, neutrino decay is a possible way out.
\section{Remarks on primordial nucleosynthesis}
As shown in the previous section, barring neutrino decay the kinematics of 
massive neutrinos after decoupling implies that the cosmic neutrino background 
as predicted from the QMF is in violent conflict with observations. 
Nevertheless, it may be of interest using the QMF to calculate the abundances 
of light nuclei synthesized in the early Universe and see if such calculations 
are consistent with observations. However, full quantitative nucleosynthesis 
calculations are beyond the scope of this paper, but some estimates and 
qualitative arguments based on the results of the previous section shall be 
made. Fortunately said calculations have already been done for power-law 
cosmologies [8,13,14], and in particular for CPLCs [14, 15]. Qualitative 
comparisons with the main results found from these calculations is the subject 
of this section.
\begin{figure}[t]
\begin{center}
\includegraphics*[height=4.8699cm,width=5.0cm,angle=0]{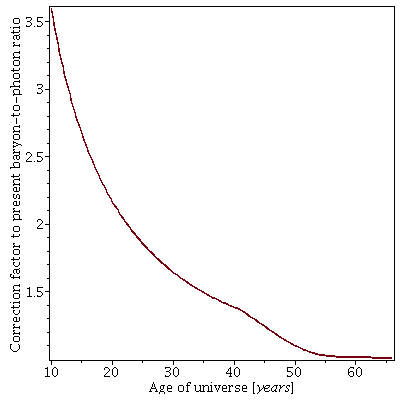}
\caption{\label{fig:corr} Correction factor to today's value of $n_b/n_{\gamma}$ 
versus age.}
\end{center}
\end{figure}

There are two main differences between primordial nucleosynthesis in the QMF 
as compared to power-law cosmologies. First, the temperature evolution for the 
power-law cosmologies is given from equation (62) for different values of 
${\alpha}$. This differs substantially from the temperature evolution given 
from equations (32) and (48). In particular, for CPLCs, the temperature
evolution goes approximately as $1/t$, whereas for the quasi-metric cosmology
we see from fig.~\ref{fig:tempkev} that the temperature drops (much) slower 
than $1/t$ during (and before) the era of electron-positron annihilation. This 
means that the quasi-metric universe at the same temperature is younger 
(and denser) than for the CPLCs during the main nucleosynthesis era. Second, 
the baryon to photon ratio $n_b/n_{\gamma}$ is not a constant in quasi-metric 
cosmology since it decreases whenever ${\beta}(T)<1$. This means that for much 
of the epoch of primordial nucleosynthesis, the value of $n_b/n_{\gamma}$ was 
{\em higher} than it is today (see fig.~\ref{fig:corr}). On the other hand, 
for the power-law cosmologies (and for standard cosmology), said value is 
constant from the epoch of primordial nucleosynthesis until today.

To get primordial nucleosynthesis started, there must be a supply of neutrons.
For a CPLC as for quasi-metric cosmology, the universe is so old at the 
relevant temperatures that the only way to have a supply of neutrons
is to produce them is via the weak-interaction reactions [15]
\eqa
n{\rightleftharpoons}p+e^-+{\bar {\nu}}_e, \qquad n+e^+{\rightleftharpoons}p+
{\bar {\nu}}_e, \qquad p+e^-{\rightleftharpoons}n+{\nu}_e.
\ena
Due to slowly changing plasma temperatures, these reactions will remain in 
equilibrium almost down to temperatures where significant neutrino decoupling 
occurs. See [16] for explict expressions for the reaction rates (per nucleon)
${\lambda}_{{\rm n}{\rightarrow}{\rm p}}$ and ${\lambda}_{{\rm p}{\rightarrow}{\rm n}}$ for
the reactions shown in equation (65). (Note that in general,  $T_{\nu}{\neq}
T_{\gamma}$ in said expressions.) Now the period where the $^4$He is synthesized 
is limited to the period before the reactions $p{\rightarrow}n$ shown in 
equation (65) freeze out, since after freeze-out, the neutron-to-proton ratio 
will no longer depend on temperature and subsequently the free neutrons will 
decay quickly leaving almost no neutrons left available for additional 
nucleosynthesis. Said limitation of the $^4$He-synthesis period is valid both 
for the CPLCs and for the quasi-metric universe, but as mentioned above the 
quasi-metric universe is younger and denser at the same temperatures but yet 
such that the temperatures drop more slowly. Also $n_b/n_{\gamma}$ is higher at 
early epochs than today. All this means that there is more time available for 
$^4$He-production at the relevant temperatures than in the CPLCs, so it might 
be possible to produce a sufficient amount of $^4$He without needing a very 
high ratio $n_b/n_{\gamma}$ today. On the other hand, to get the right amount of 
$^4$He for a CPLC (assuming that ${\mu}_{{\nu}_i}=0$), one must have 
$n_b/n_{\gamma}{\approx}1.05{\times}10^{-8}$ today [15], i.e., much larger than 
the value of $(6.10{\pm}0.04){\times}10^{-10}$ [17] as inferred by 
analysing the cosmic microwave background assuming standard cosmology.
It is possible for a CPLC to produce a sufficient amount of $^4$He with a 
significantly smaller value of $n_b/n_{\gamma}$ by assuming a suitable neutrino 
asymmetry (i.e., ${\mu}_{{\nu}_i}<0$) [15].

Another serious problem for both the CPLCs and for quasi-metric cosmology
is the severe primordial underproduction of the light elements D, $^3$He and 
$^7$Li. Deuterium is produced/destroyed via the reactions
\eqa
n+p{\rightleftharpoons}{\rm D}+{\gamma}, \quad p+p{\rightarrow}{\rm D}+e^+
+{\nu}_e, \quad {\rm D}+p{\rightleftharpoons} ^3{\rm He}+{\gamma},
\quad ^3{\rm He}+{\rm D}{\rightarrow} ^4{\rm He}+p.
\ena
However, at the relevant temperatures of $^4$He-synthesis, due to the weakly
bound D-nucleus and the long time available for nucleosynthesis, almost any 
produced D has been destroyed when the reactions shown in equation (66) 
eventually freeze out. The same comment applies to most produced $^3$He via
the third reaction shown in equation (66), since it will be destroyed via the
reaction $^3{\rm He}+p{\rightarrow}^4{\rm He}+e^++{\nu}_e$. Similarly, almost 
any produced $^7$Li will be destroyed via the reaction
$^7{\rm Li}+p{\rightarrow}^4{\rm He}+^4{\rm He}$. See, e.g., [14, 15] for
detailed abundances calculated as functions of temperature for CPLCs. To get 
acceptable levels of D, $^3$He and $^7$Li, these nuclei must be produced much
later in the history of the universe, in environments where destruction rates 
of said nuclei are not crucial. For the CPLCs, spallation processes connected 
to protostar formation have been proposed as a mechanism for producing the 
missing light elements [15].

Quasi-metric cosmology shares the problem of said missing light elements with
the CPLCs. However, since the primordial abundance of D is inferred from
analysing absorption line systems in spectra of quasi-stellar objects (QSOs)
(D/H$=(2.53{\pm}0.10){\times}10^{-5}$ by number [18]), the nature of QSOs may be
crucial when interpreting these results. That is, the interpretation of QSOs 
and their absorption line systems within quasi-metric cosmology may be very 
different from the standard one and such that there is a possibility that the 
missing light elements may be produced within the QSOs themselves (possibly 
via spallation mechanisms). Moreover, there are in fact no direct observational
data on primordial abundances of $^3$He; data for abundances of $^3$He
($^3$He/H$=(1.01{\pm}0.2){\times}10^{-5}$ by number [19]) are limited to within 
our galaxy and represent upper limits to the primordial abundance.
\section{Conclusion}
The main result of this paper is that assuming the validity of standard 
neutrino physics and the approximate model given by equation (42),
the calculated properties of the predicted cosmic relic
neutrino background as obtained from the QMF are in gross conflict with 
observations. This means that, with no natural explanation to resolve said 
conflict, the QMF is currently nonviable. The nonviable status of the QMF can 
only be revoked by experimental evidence indicating new, exotic neutrino 
physics. One theoretical possibility of such is if the lightest mass eigenstate
${\nu}_{1}$ (normal hierarchy) is null (and thus stable) and the other two 
massive eigenstates decay into ${\nu}_{1}$ and some other (possibly massless) 
particle $X$, i.e.,
\eqa
{\nu}_{i}{\rightarrow}{\nu}_{1}+X, \qquad i{\in}{\{}2,3{\}}.
\ena
Since there are strong restrictions on neutrino radiative decay from 
observations of solar neutrinos and neutrinos observed from supernova 1987A
(see, e.g., [20]), $X$ can hardly be a photon. That is, said restrictions yield
${\tau}_{{\nu}_i,0}/m_{{\nu}_i}c^2{\gtrsim}10^{15}$ s/eV, where ${\tau}_{{\nu}_i,0}$ is 
the lifetime of mass eigenstate ${\nu}_{i}$ in its rest frame. However, the 
reduction in relic neutrino numbers with a factor 1000 or more needed to avoid 
said conflict with experiment yields 
${\tau}_{{\nu}_i,0}/m_{{\nu}_i}c^2{\lesssim}10^{11}$ s/eV (by solving the decay 
equation). On the other hand, neutrino non-radiative decay is much less 
restricted by observations. That is, $X$ might be a light (possibly massless) 
sterile mass eigenstate ${\nu}_{{\rm s}}$ (that may or may not be accessible 
via neutrino oscillations) [21]. We note that there exists a number of 
estimates of neutrino decay limits and lifetimes (see, e.g., [22] for a 
review), but most are model-dependent and not stringent enough to rule out a 
potential resolution of said conflict.

If future experiments show that no neutrino mass eigenstate is null, the 
only possibility left would be decay of all neutrino mass eigenstates into 
``invisible'' particles over cosmic time scales; this seems rather farfetched 
and difficult to test. Nevertheless this possibility cannot be dismissed out
of hand. However, as long as there is no independent experimental 
evidence that neutrinos really do decay, the QMF should be declared nonviable.
\\ [4mm]
{\bf Acknowledgement} \\ [1mm]
I wish to thank Dr. K{\aa}re Olaussen for reviewing the manuscript.
\\ [4mm]
{\bf References} \\ [1mm]
{\bf [1]} D. {\O}stvang, {\em Grav. \& Cosmol.} {\bf 11}, 205 (2005)
(gr-qc/0112025).  \\
{\bf [2]} D. {\O}stvang, {\em Doctoral thesis}, (2001) (gr-qc/0111110). \\
{\bf [3]} D. {\O}stvang, {\em Grav. \& Cosmol.} {\bf 13}, 1 (2007)
(gr-qc/0201097). \\
{\bf [4]} V. Mukhanov, {\em Physical Foundations of Cosmology}, \\
{\hspace*{6.3mm}} Cambridge University Press, 2005. \\
{\bf [5]} E.W. Kolb and M.S. Turner, {\em The Early Universe}, 
Addison-Wesley, 1990. \\
{\bf [6]} C. Giunti and C.W. Kim, {\em Fundamentals of Neutrino Physics 
and Astrophysics}, \\
{\hspace*{5.1mm}} Oxford University Press, 2007. \\
{\bf [7]} D.A. Dicus, {\em Phys. Rev.} {\bf D 6}, 941 (1972). \\
{\bf [8]} A. Batra, D. Lohiya, S. Mahajan and A. Mukherjee, 
{\em IJMP} {\bf D 9}, 757 (2000) \\
{\hspace*{7.1mm}}(nucl-th/9902022). \\
{\bf [9]} H. Zhang, {\em Class. Quantum Grav.} {\bf 25}, 208001 (2008)
(arXiv:0808.1552) \\
{\bf [10]} M.C. Gonzalez-Garcia, M. Maltoni, J. Salvado and T. Schwetz,
{\em JHEP} {\bf 12}, 123 (2012) \\
{\hspace*{7.7mm}}(arXiv:1209.3023). \\
{\bf [11]} G. Bellini {\em et al.}, {\em Nature} {\bf 512}, 383 (2014). \\ 
{\bf [12]} J.N. Abdurashitov {\em et al.},
{\em Phys. Rev.} {\bf C 80}, 015807 (2009) (arXiv:0901.2200). \\
{\bf [13]} M. Sethi, A. Batra and D. Lohiya, {\em Phys. Rev.} {\bf D 60}, 
108301 (1999) \\
{\hspace*{7.7mm}}(astro-ph/9903084). \\
{\bf [14]} M. Kaplinghat, G. Steigman and T. P. Walker, 
{\em Phys. Rev.} {\bf D 61}, 103507 (2000) \\
{\hspace*{7.7mm}}(astro-ph/9911066). \\
{\bf [15]} P. Singh and D. Lohiya, {\em JCAP} {\bf 05}, 
061 (2015) (arXiv:1312.7706). \\
{\bf [16]} G. Beaudet and P. Goret, {\em Astron. Astrophys.} {\bf 49}, 415 
(1976). \\
{\bf [17]} P.A.R. Ade {\em et al.}, {\em A{\&}A} {\bf 594}, A13
(2016) (arXiv:1502.01589). \\
{\bf [18]} R. Cooke {\em et al.}, {\em ApJ} {\bf 781}, 31 (2014). \\
{\bf [19]} T.M. Bania, R.T. Rood and D. S. Balser, 
{\em Nature} {\bf 415}, 54 (2002). \\
{\bf [20]} E.L. Chupp, W.T. Vestrand and C. Reppin, {\em Phys. Rev. Lett.} 
{\bf 62}, 505 (1989). \\
{\bf [21]} J.F. Beacom and N.F. Bell, {\em Phys. Rev.} {\bf D 65}, 113009 
(2002) (hep-ph/0204111). \\
{\bf [22]} K. A. Olive {\em et al.}, {\em Chin. Phys.} {\bf C 38}, 
090001 (2014).  \\
{\bf [23]} F. Debbasch and W.A. van Leeuwen, {\em Physica} {\bf A 388}, 1818
(2009). \\
\appendix
\renewcommand{\theequation}{\thesection.\arabic{equation}}
\setcounter{equation}{0}
\section{Integrodifferential equations}
In this appendix we give explicit expressions for the integrodifferential 
equations (32) for $T(t)$ and (45) for $h(t)$, using equation (48) for 
${\beta}(T)$. In addition we have the thermally non-averaged decoupling 
criterion (50) for $E_{\rm D}$. Setting $w{\equiv}E_{\rm D}/m_ec^2$, 
$y{\equiv}m_ec^2/k_{\rm B}T$, equation (32) reads
\eqa
t{\frac{dy}{dt}}-{\beta}(y,w,h,t)y=0,
\ena
where ${\beta}(y,w,h,t)$ is found from equation (48), i.e.,
\eqa
{\beta}(y,w,h,t){\approx}
{\frac{1+{\frac{45y^4}{2{\pi}^4}}
{\int}_{{\!}{\!}1}^{\infty}
{\frac{x^2{\sqrt{x^2-1}}dx}{1+{\exp}(yx)}}}
{1+{\frac{30y^4}{{\pi}^4}}{\int}_{{\!}{\!}1}^{\infty}
{\frac{x^2{\sqrt{x^2-1}}dx}{1+{\exp}(yx)}}
+{\frac{15y^4}{2{\pi}^4}}{\int}_{{\!}{\!}1}^{\infty}
{\frac{x^2dx}{{\sqrt{x^2-1}}[1+{\exp}(yx)]}}+{\frac{45h}{4{\pi}^4}}
{\int}_{{\!}{\!}yw}^{\infty}{\frac{x^3dx}{1+{\exp}(x)}}}} \nonumber \\
+{\frac{{\frac{45h}{4{\pi}^4}}{\int}_{{\!}{\!}yw}^{\infty}
{\frac{x^3dx}{1+{\exp}(x)}}
[{\Gamma}^{\rm net}_{\rm ann}-{\frac{2}{3{\zeta}(3)}}
{\int}_{{\!}{\!}0}^{yw}{\frac{x^2dx}{1+{\exp}(x)}}
{\Gamma}_{\rm scat}^{\rm eff}]t 
+{\frac{15h}{2{\pi}^4{\zeta}(3)}}
{\int}_{{\!}{\!}0}^{yw}{\frac{x^3dx}{1+{\exp}(x)}}
{\int}_{{\!}{\!}yw}^{\infty}{\frac{x^2dx}{1+{\exp}(x)}}
{\Gamma}_{\rm scat}^{\rm eff}t}
{1+{\frac{30y^4}{{\pi}^4}}{\int}_{{\!}{\!}1}^{\infty}
{\frac{x^2{\sqrt{x^2-1}}dx}{1+{\exp}(yx)}}
+{\frac{15y^4}{2{\pi}^4}}{\int}_{{\!}{\!}1}^{\infty}
{\frac{x^2dx}{{\sqrt{x^2-1}}[1+{\exp}(yx)]}}+{\frac{45h}{4{\pi}^4}}
{\int}_{{\!}{\!}yw}^{\infty}{\frac{x^3dx}{1+{\exp}(x)}}}}.
\ena
Here, we have from equations (49) and (60) that
\eqa
{\Gamma}_{\rm scat}^{\rm eff}={\frac{3}{2}}{\zeta}(3)
K{\frac{I(y)}{{\frac{3}{2}}{\zeta}(3)-J(yw)}}
{\Big {\{}}{\frac{b}{y}}{\Big [}I_1(y,w)-I_3(y,w)+
{\frac{2}{3}}I_4(y,w){\Big ]}-aI_2(y,w){\Big {\}}},
\ena
where we have defined $a{\equiv}
2{\sin}^2{\theta}_{\rm w}(6{\sin}^2{\theta}_{\rm w}-1)
{\approx}0.1748$, $b{\equiv}24{\sin}^4{\theta}_{\rm w}-
4{\sin}^2{\theta}_{\rm w}+3{\approx}3.3496$ and $K{\equiv}
{\frac{4G_{\rm F}^2m_e^5c^4}{9{\pi}^3{\zeta}(3){\hbar}^7}}{\approx}
8.5855{\times}10^{-5}$ s$^{-1}{\approx}2.7075{\times}10^{3}$ yr$^{-1}$. 
Also, $I_1(y,w)$, $I_2(y,w)$, $I_3(y,w)$ and $I_4(y,w)$  are given 
from equation (57) and we have defined 
\eqa
I(y){\equiv}{\int}_{{\!}{\!}{\!}{\!}1}^{\infty}
{\frac{x{\sqrt{x^2-1}}dx}{1+{\exp}(yx)}},
\ena
in addition to
\eqa
J(u){\equiv}{\int}_{{\!}{\!}{\!}{\!}0}^u{\frac{x^2dx}{1+e^x}}=
{\frac{3}{2}}{\zeta}(3)
+{\frac{1}{3}}u^3-u^2{\ln}(1+e^u)-2u{\rm Li}_2(-e^u)+2{\rm Li}_3(-e^u),
\ena
where ${\rm Li}_s(z)$ is the polylogarithm, defined by
\eqa
{\rm Li}_s(z){\equiv} \sum_{k=1}^{\infty}{\frac{z^k}{k^s}}, \qquad 
z{\in}{\bf C}.
\ena
We notice the exact expression
\eqa
{\int}_{{\!}{\!}{\!}{\!}u}^{\infty}{\frac{x^3dx}{1+e^x}}=
{\frac{7{\pi}^4}{60}}-{\frac{1}{4}}u^4+u^3{\ln}(1+e^u)
+3u^2{\rm Li}_2(-e^u)-6u{\rm Li}_3(-e^u)+6{\rm Li}_4(-e^u).
\ena
Next, for epochs $t{\leq}t_ {\rm C}$, we have from equation (45) that
\eqa
{\Gamma}^{\rm net}_{\rm ann}{\approx}{\frac{3}{t}}{\Big (}1-{\beta}(y,w,t)
{\Big )}+{\frac{2}{3{\zeta}(3)}} \int_0^{yw}{\frac{x^2dx}
{{\exp}(x)+1}}{\Gamma}^{\rm eff}_{\rm scat}, \qquad t{\leq}t_{\rm C},
\ena
since $h=1$ and ${\dot h}=0$ for this time interval. For later times 
$t{\geq}t_{\rm C}$, ${\Gamma}^{\rm net}_{\rm ann}$ will be smaller and its decline 
is determined by the evolution of the thermally averaged annihilation rate 
(producing neutrinos). The latter can be found from the number density of 
electrons/positrons and the size of the thermally averaged cross sections 
shown in equation (61). Therefore, with 
${\beta}_{\rm C}{\equiv}{\beta}(y_{\rm C},w_{\rm C},t_{\rm C})$,
$I_{1{\rm C}}{\equiv}I_1(y_{\rm C},w_{\rm C})$ etc., we can set (see also equation 
(A.18) below)
\eqa
{\Gamma}^{\rm net}_{\rm ann}{\approx}{\Big [}{\frac{3}{t_{\rm C}}}
(1-{\beta}_{\rm C})+{\frac{2}{3{\zeta}(3)}}
J(y_{\rm C}w_{\rm C}){\Gamma}^{\rm eff}_{\rm scat}(y_{\rm C},w_{\rm C}){\Big ]}
\nonumber \\
{\times}{\frac{I(y){\Big (}b[4I_5(y)-I(y)]+3a[2I(y)-{\frac{1}{2}}I_6(y)]
{\Big )}}{I(y_{\rm C}){\Big (}b[4I_5(y_{\rm C})-I(y_{\rm C})]
+3a[2I(y_{\rm C})-{\frac{1}{2}}I_6(y_{\rm C})]{\Big )}}}
\nonumber \\
={\Big [}{\frac{3}{t_{\rm C}}}(1-{\beta}_{\rm C})
+{\frac{KJ({y_{\rm C}w_{\rm C}})I(y_{\rm C})}{{\frac{3}{2}}{\zeta}(3)-
J({y_{\rm C}w_{\rm C}})}}{\Big {\{}}{\frac{b}{y_{\rm C}}}{\Big (}
I_{1{\rm C}}-I_{3{\rm C}}+{\frac{2}{3}}I_{4{\rm C}}{\Big )}-
aI_{2{\rm C}}{\Big {\}}}{\Big ]} \nonumber \\
{\times}{\frac{I(y){\Big (}b[4I_5(y)-I(y)]+3a[2I(y)-{\frac{1}{2}}I_6(y)]
{\Big )}}{I(y_{\rm C}){\Big (}b[4I_5(y_{\rm C})-I(y_{\rm C})]
+3a[2I(y_{\rm C})-{\frac{1}{2}}I_6(y_{\rm C})]{\Big )}}}, \qquad 
t{\geq}t_{\rm C},
\ena
where $I_5(y)$ and $I_6(y)$ are defined in equation (59).
Now, combining equations (45) and (A.9) we find the equation
\eqa
{\frac{1}{h}}{\frac{dh}{dt}}+{\frac{3}{t}}(1-{\beta})+
{\frac{KJ(yw)I(y)}{{\frac{3}{2}}{\zeta}(3)-J(yw)}} {\Big {\{}}
{\frac{b}{y}}{\Big [}I_1-I_3+
{\frac{2}{3}}I_4{\Big ]}-aI_2{\Big {\}}} \nonumber \\
-{\Big [}{\frac{3}{t_{\rm C}}}(1-{\beta}_{\rm C})
+{\frac{KJ(y_{\rm C}w_{\rm C})I(y_{\rm C})}{{\frac{3}{2}}{\zeta}(3)-J
(y_{\rm C}w_{\rm C})}}
{\Big {\{}}{\frac{b}{y_{\rm C}}}{\Big [}I_{1{\rm C}}-I_{3{\rm C}}+
{\frac{2}{3}}I_{4{\rm C}}{\Big ]}-aI_{2{\rm C}}{\Big {\}}}{\Big ]}
\nonumber \\
{\times}{\frac{I(y){\Big (}b[4I_5(y)-I(y)]+3a[2I(y)-{\frac{1}{2}}I_6(y)]
{\Big )}}{I(y_{\rm C}){\Big (}b[4I_5(y_{\rm C})-I(y_{\rm C})]
+3a[2I(y_{\rm C})-{\frac{1}{2}}I_6(y_{\rm C})]{\Big )}}}=0, \qquad 
t{\geq}t_{\rm C},
\ena
valid for $t{\geq}t_{\rm C}$. The third equation needed to close the set of 
equations is obtained from the criterion (51), written in the form
\eqa
G_{\rm scat}(y,w,h,t)={\frac{3}{2}}{\zeta}(3)K
{\frac{tI(y)}{{\beta}(y,w,h,t)}}
{\frac{w^2}{1+2w}}{\Big [}b-{\frac{a+bw}{1+2w}}+
{\frac{2bw^2}{3(1+2w)^2}}{\Big ]}{\approx}1.
\ena 
However, in practice it may be easier to solve numerically a third 
integrodifferential equation rather than the implicit relationship (A.11). In 
particular, trying to solve the coupled set of equations (A.1), (A.10) and 
(A.11) using MAPLE does not work since equation (A.11) is not separable in the 
variable $w$. This problem can be circumvented by taking the derivative of 
(A.11) with respect to $t$. We then get the equation
\eqa
{\frac{3}{2}}{\zeta}(3)K{\frac{w^2}{1+2w}}{\Big [}b-{\frac{a+bw}{1+2w}}+
{\frac{2bw^2}{3(1+2w)^2}}{\Big ]}{\Big {\{}}I(y)-
{\int}_{{\!}{\!}{\!}{\!}1}^{\infty}{\frac{e^{yx}x^2{\sqrt{x^2-1}}dx}
{[1+{\exp}(yx)]^2}}{\beta}y{\Big {\}}}
\nonumber \\
+{\frac{1}{2}}{\zeta}(3)KI(y)
{\frac{w}{(1+2w)^4}}[16bw^3+32bw^2-(12a-21b)w-6(a-b)]
{\frac{tdw}{dt}}{\approx}{\frac{d{\beta}}{dt}},
\ena
where the right hand side can be found from equation (A.2), i.e.,
\eqa
{\Big {\{}}1+{\frac{30y^4}{{\pi}^4}}{\int}_{{\!}{\!}{\!}{\!}1}^{\infty}
{\frac{x^2{\sqrt{x^2-1}}dx}{1+e^{yx}}} 
+{\frac{15y^4}{2{\pi}^4}}{\int}_{{\!}{\!}{\!}{\!}1}^{\infty}
{\frac{x^2dx}{{\sqrt{x^2-1}}[1+e^{yx}]}}+{\frac{45h}{4{\pi}^4}}
{\int}_{{\!}{\!}{\!}{\!}yw}^{\infty}{\frac{x^3dx}{1+e^x}}{\Big {\}}}
{\frac{d{\beta}}{dt}}
\nonumber \\
={\Big [}{\frac{90y^3}{{\pi}^4}}{\int}_{{\!}{\!}{\!}{\!}1}^{\infty}
{\frac{x^2{\sqrt{x^2-1}}dx}{1+{\exp}(yx)}}-
{\frac{45y^4}{2{\pi}^4}}{\int}_{{\!}{\!}{\!}{\!}1}^{\infty}
{\frac{e^{yx}x^3{\sqrt{x^2-1}}dx}{[1+{\exp}(yx)]^2}}
-{\frac{45h}{4{\pi}^4}}{\frac{y^3w^4}{1+{\exp}(yw)}}
\nonumber \\
{\times}{\Big (}{\Gamma}^{\rm net}_{\rm ann}-{\Gamma}^{\rm eff}_{\rm scat}
{\Big )}t+{\beta}{\Big {\{}}-{\frac{120y^3}{{\pi}^4}}
{\int}_{{\!}{\!}{\!}{\!}1}^{\infty}
{\frac{x^2{\sqrt{x^2-1}}dx}{1+e^{yx}}}+{\frac{30y^4}{{\pi}^4}}
{\int}_{{\!}{\!}{\!}{\!}1}^{\infty}
{\frac{e^{yx}x^3{\sqrt{x^2-1}}dx}{[1+e^{yx}]^2}}
\nonumber \\
-{\frac{30y^3}{{\pi}^4}}{\int}_{{\!}{\!}{\!}{\!}1}^{\infty}
{\frac{x^2dx}{{\sqrt{x^2-1}}[1+e^{yx}]}}
+{\frac{15y^4}{2{\pi}^4}}{\int}_{{\!}{\!}{\!}{\!}1}^{\infty}
{\frac{e^{yx}x^3dx}{{\sqrt{x^2-1}}[1+e^{yx}]^2}}+
{\frac{45h}{4{\pi}^4}}{\frac{y^3w^4}{1+e^{yw}}}{\Big {\}}}
{\Big ]}{\frac{{\beta}y}{t}}
\nonumber \\
+{\frac{45h}{4{\pi}^4}}{\frac{y^4w^3}{1+e^{yw}}}{\Big (}
{\beta}-[{\Gamma}^{\rm net}_{\rm ann}-{\Gamma}^{\rm eff}_{\rm scat}]t
{\Big )}{\frac{dw}{dt}}-
{\frac{7h}{16{\zeta}(3)}}{\frac{y^3w^2}{1+e^{yw}}}
{\Gamma}^{\rm eff}_{\rm scat}{\Big (}{\beta}w+t{\frac{dw}{dt}}{\Big )}
\nonumber \\
+{\frac{45}{4{\pi}^4}}{\int}_{{\!}{\!}{\!}{\!}yw}^{\infty}{\frac{x^3dx}{1+e^x}}
{\Big (}[{\Gamma}^{\rm net}_{\rm ann}-{\Gamma}^{\rm eff}_{\rm scat}]t
-{\beta}{\Big )}{\frac{dh}{dt}}+{\frac{7}{16}}{\Big(}{\frac{3}{2}}-
{\frac{J(yw)}{{\zeta}(3)}}{\Big )}{\Gamma}^{\rm eff}_{\rm scat}
{\Big (}h+t{\frac{dh}{dt}}{\Big )}
\nonumber \\
+{\frac{45h}{4{\pi}^4}}{\int}_{{\!}{\!}{\!}{\!}yw}^{\infty}
{\frac{x^3dx}{1+e^x}}{\Big (}{\Gamma}^{\rm net}_{\rm ann}-
{\Gamma}^{\rm eff}_{\rm scat}+t{\frac{d}{dt}}{\Gamma}^{\rm net}_{\rm ann}
-t{\frac{d}{dt}}{\Gamma}^{\rm eff}_{\rm scat}{\Big )}
+{\frac{7h}{16}}{\Big(}{\frac{3}{2}}-{\frac{J(yw)}{{\zeta}(3)}}{\Big )}
t{\frac{d}{dt}}{\Gamma}^{\rm eff}_{\rm scat}.
\ena
Here we have
\eqa
{\frac{d}{dt}}{\Gamma}^{\rm eff}_{\rm scat}= 
{\Big [}{\frac{y^3w^2({\frac{{\beta}w}{t}}+{\frac{dw}{dt}})}
{[{\frac{3}{2}}{\zeta}(3)-J(yw)][1+e^{yw}]}}
-{\frac{{\beta}}{t}}{\Big (}1+{\frac{y}{I(y)}}
{\int}_{{\!}{\!}{\!}{\!}1}^{\infty}
{\frac{e^{yx}x^2{\sqrt{x^2-1}}dx}{[1+e^{yx}]^2}}{\Big )} \nonumber \\
-{\frac{{\frac{{\beta}}{t}}{\{}(a+b)I_2
-2b{\int}_{{\!}{\!}yw}^{\infty}{\frac{x^5dx}{[y+2x]^3[1+e^x]}}
+2b{\int}_{{\!}{\!}yw}^{\infty}{\frac{x^6dx}{[y+2x]^4[1+e^x]}}
-2ay{\int}_{{\!}{\!}yw}^{\infty}{\frac{x^4dx}{[y+2x]^3[1+e^x]}}
{\}}}
{{\frac{b}{y}}(I_1-I_3+{\frac{2}{3}}I_4)-aI_2}} \nonumber \\
+{\frac{({\frac{{\beta}w}{t}}+{\frac{dw}{dt}})y^3w^4
(-b+{\frac{a+bw}{1+2w}}-{\frac{2}{3}}{\frac{bw^2}{[1+2w]^2}})}
{[1+2w][1+e^{yw}][{\frac{b}{y}}(I_1-I_3+{\frac{2}{3}}I_4)-aI_2]}}{\Big ]}
{\Gamma}^{\rm eff}_{\rm scat},
\ena
\eqa
{\frac{d}{dt}}{\Gamma}^{\rm net}_{\rm ann}{\approx}-{\frac{3}{t^2}}(1-{\beta})
-{\frac{3}{t}}{\frac{d{\beta}}{dt}}+
{\frac{2}{3{\zeta}(3)}}{\Big [}{\frac{y^3w^2
({\frac{{\beta}w}{t}}+{\frac{dw}{dt}})}{1+e^{yw}}}{\Gamma}^{\rm eff}_{\rm scat}
+J(yw){\frac{d}{dt}}{\Gamma}^{\rm eff}_{\rm scat}{\Big ]},
\ena
for $t<t_{\rm C}$, and
\eqa
{\frac{d}{dt}}{\Gamma}^{\rm net}_{\rm ann}{\approx}-
{\Big {\{}}{\frac{1}{I(y)}}{\int}_{{\!}{\!}{\!}{\!}1}^{\infty}
{\frac{e^{yx}x^2{\sqrt{x^2-1}}dx}{[1+{\exp}(yx)]^2}} 
+{\frac{b[4{\int}_{{\!}{\!}1}^{\infty}{\frac{e^{yx}x^4
{\sqrt{x^2-1}}dx}{[1+{\exp}(yx)]^2}}-{\int}_{{\!}{\!}1}^{\infty}
{\frac{e^{yx}x^2{\sqrt{x^2-1}}dx}{[1+{\exp}(yx)]^2}}]}
{b[4I_5(y)-I(y)]+3a[2I(y)-{\frac{1}{2}}I_6(y)]}}
\nonumber \\
+{\frac{3a[2{\int}_{{\!}{\!}1}^{\infty}
{\frac{e^{yx}x^2{\sqrt{x^2-1}}dx}{[1+{\exp}(yx)]^2}}
-{\frac{1}{2}}{\int}_{{\!}{\!}1}^{\infty}
{\frac{e^{yx}{\sqrt{x^2-1}}dx}{[1+{\exp}(yx)]^2}}]}
{b[4I_5(y)-I(y)]+3a[2I(y)-{\frac{1}{2}}I_6(y)]}}
{\Big {\}}}{\Gamma}^{\rm net}_{\rm ann}{\frac{{\beta}y}{t}}, \qquad 
t{\geq}t_{\rm C}.
\ena
We note that equations (A.1), (A.10) and (A.12) are integrodifferential 
equations rather than ordinary differential equations which represents 
numerical challenges. Therefore it is convenient to approximate these equations
with ordinary differential equations. To do that, it is necessary to substitute
the integrals with the factor $[1+{\exp}(yx)]^{-1}$ in the integrand with 
approximate integrals obtained by making the approximation $1+{\exp}(yx)
{\approx}{\exp}(yx)$. Numerically it is found that this overestimates the 
integrals by a negligible amount for the relevant range of $y{\gg}1$. The 
approximate integrals can be expressed by modified Bessel functions recognized 
by MAPLE. Besides, in equations (A.10) and (A.12), for the integrals having the 
factor $[1+{\exp}(x)]^{-1}$ in the integrand, the approximation $1+
{\exp}(x){\approx}{\exp}(x)$ can be made, and the errors so introduced will 
not exceed about 5${\%}$ for $T<T_{\rm C}$ (for $T>T_{\rm C}$ the errors will be
larger, but will not significantly change the results). These approximate 
integrals can be also be expressed by special functions recognized by MAPLE. 
In this way the integrodifferential equations can be well approximated by 
ordinary differential equations, and the coupled set of equations (A.1), (A.10) 
and (A.12) can be solved numerically using MAPLE. (For $t<t_{\rm C}$, equation 
(A.10) becomes redundant, of course.)

To solve said set of approximative differential equations, one must first guess
the critical temperature $T_{\rm C}$ (or equivalently, $y_{\rm C}$) used as a 
boundary value. The criteria (50) and (51) then yield two coupled equations for
the corresponding time $t_{\rm C}$ and $w_{\rm C}{\equiv}w(t_{\rm C})$. Here we use
that $h_{\rm C}{\equiv}h(t_{\rm C})=1$ while the criterion (51) yields (A.11). The 
criterion (50) then reads (here $I_{\rm C}{\equiv}I(y_{\rm C})$ and
$J_{\rm C}{\equiv}J(y_{\rm C}w_{\rm C})$)
\eqa
G_{\rm ann}{\approx}{\frac{{\frac{{\zeta}(3)}{4}}Ky_{\rm C}^3I_{\rm C}
[4bI_{5{\rm C}}-(b-6a)I_{\rm C}-{\frac{3a}{2}}I_{6{\rm C}}]}
{[{\frac{3}{t_{\rm C}}}(1-{\beta}_{\rm C})]
[{\frac{3}{2}}{\zeta}(3)-J_{\rm C}]
+KJ_{\rm C}I_{\rm C}[{\frac{b}{y_{\rm C}}}(I_{1{\rm C}}-I_{3{\rm C}}+
{\frac{2}{3}}I_{4{\rm C}})-aI_{2{\rm C}}]
}}{\approx}1,
\ena
and inserted into equation (A.8) this yields
\eqa
{\Gamma}^{\rm net}_{\rm ann}{\approx}
{\frac{{\zeta}(3)Ky_{\rm C}^3I(y){\{}b[4I_5(y)-I(y)]+3a[2I(y)-
{\frac{1}{2}}I_6(y)]{\}}}{
4[{\frac{3}{2}}{\zeta}(3)-J(y_{\rm C}w_{\rm C})]}}, \qquad t{\geq}t_{\rm C}.
\ena
Note that we may safely guess $w_{\rm C}{\approx}0$ when solving (A.11) and 
(A.17) numerically to determine $w_{\rm C}$ and $t_{\rm C}$ for a given $y_{\rm C}$.

Using the boundary conditions $y_{\rm C}$, $h_{\rm C}=1$ and
$w_{\rm C}$, said coupled set of differential equations can be solved until
the decoupling temperature $T_{\rm dec}$ (or equivalently, $y_{\rm dec}$) is
reached. At this temperature, the contribution to ${\beta}(y,h,w,t)$ from 
neutrinos is negligible. However, we must also solve equation (A.1) for 
$t>t_{\rm dec}$, using the values $t_0{\approx}H_0^{-1}{\approx}1.34{\ }
{\times}10^{10}{\rm yr}$ (for the choice $H(t_0)=73{\ }{\frac{{\rm km}}
{{\rm sMpc}}}{\approx}2.3658{\ }{\times}10^{-18}{\rm s}^{-1}$) and 
$T_0{\approx}2.725$ K (or equivalently, $y_0{\approx}2.1726{\times}10^9$) as 
boundary values for the present era. To be consistent, this calculation should 
agree with the result $y_{\rm dec}(t_{\rm dec})$. If the two results do not match, 
a different choice of $y_{\rm C}$ must be made until a value is found that 
yields agreement between the two solutions.

Once the functions $y(t)$, $w(t)$ and $h(t)$ have been found numerically, 
equation (44) can also be solved numerically with said solutions declared as 
known functions in MAPLE. Equation (44) may conveniently be written in the form
\eqa
{\frac{du}{dt}}+{\frac{3}{t}}u-
KJ(yw)I(y){\Big {\{}}{\frac{b}{y}}{\Big [}I_1-I_3+
{\frac{2}{3}}I_4{\Big ]}-aI_2{\Big {\}}}{\frac{h}{y^3}}-
{\frac{hw^2}{y(1+e^{yw})}}{\frac{d}{dt}}(yw)=0,
\ena
where $u{\equiv}n^{\rm dec}_{\nu}/K_{\rm nd}$ and $K_{\rm nd}{\equiv}
{\frac{m_e^3c^3}{2{\pi}^2{\hbar}^3}}{\approx}8.7977{\times}10^{29}$ cm$^{-3}$.
Equation (A.19) may be solved by using the initial condition $u=0$ for some
conveniently chosen early epoch. (To be able to solve equation (A.19) using 
MAPLE, it is of course again necessary to approximate the integrals entering 
it with functions recognized by MAPLE.)
\setcounter{equation}{0}
\section{Quasi-metric Boltzmann equations}
It would seem reasonable to assume that the state of matter in the early 
universe can be approximately described as a plasma in thermal equilibrium. But
this is only true as long as no processes force significant deviations from it,
a condition that will not hold in general. On the contrary, important processes
such as neutrino decoupling and primordial nucleosynthesis are examples of 
physical effects that may involve significant departures from thermal 
equilibrium.

To properly describe the thermodynamical state of matter if it deviates 
significantly from equilibrium, one should calculate the evolution of the 
particles' phase-space distribution function $f(t,x^{\mu},{\bf p}_{t})$ (where 
${\bf p}_t$ is the particle 4-momentum) with cosmic epoch. To accomplish this 
means that one needs a quasi-metric counterpart to the general-relativistic 
Boltzmann equation. A thorough treatment of the general-relativistic Boltzmann
equation is given in [23], and a quasi-metric counterpart to it can be found 
just by including the extra effects of the cosmic expansion via the global time 
function $t$, being consistent with the equations of motion (6). The main
complication is that in quasi-metric space-time, the cosmic expansion affects
photons and material particles differently, meaning that two separate equations
must be found. The quasi-metric Boltzmann equation {\em for material particles}
is closest in form to the general-relativistic case since there will be no 
``force'' term due to the cosmic expansion. Following the terminology in [23]
we find
\eqa
{\frac{dt}{d{\tau}_t}}{\frac{{\partial}f}{{\partial}t}}+
{\frac{dx^{\mu}}{d{\tau}_t}}{\frac{{\partial}f}{{\partial}x^{\mu}}}-
{\topstar{\Gamma}}_{{\beta}{\nu}}^{{\,}i}{\frac{dx^{\beta}}{d{\tau}_t}}
p_{(t)}^{\nu}{\frac{{\partial}f}{{\partial}p_{(t)}^i}}=C_{\rm coll}[f],
\ena
where ${\tau}_t$ is the proper time along the particle path and where
the momentum components $p_{(t)0}$ and $p_{(t)}^{\mu}$ should be taken as functions
of the independent variables $t$, $x^0$ and $p_{(t)i}$ [23]. Moreover, 
$C_{\rm coll}[f]$ is the collision term describing the effects of all relevant
collisional interactions, given by [23]
\eqa
C_{\rm coll}[f]{\equiv}{\int}{\int}{\int}[f({p'}_{(t)}^{i})f({q'}_{(t)}^{i})
{\hat w}({p'}_{(t)}^{i},{q'}_{(t)}^{i}{\mid}p_{(t)}^i,q_{(t)}^i) \nonumber \\
-f(p_{(t)}^i)f(q_{(t)}^i){\hat w}(p_{(t)}^i,q_{(t)}^i{\mid}
{p'}_{(t)}^{i},{q'}_{(t)}^{i})]d^3V_{q_t}d^3V_{q'_t}d^3V_{p'_t},
\ena
where the integrations are done over momentum spaces. Here, 
${\hat w}({p'}_{(t)}^{i},{q'}_{(t)}^{i}{\mid}p_{(t)}^i,q_{(t)}^i)$ is a transition 
rate, i.e., a measure of the probability that a collision between two particles
with initial momenta ${\bf p}'_t$ and ${\bf q}'_t$ and final momenta 
${\bf p}_t$ and ${\bf q}_t$ will occur [23].

On the other hand, the Boltzmann equation {\em for photons} (or null particles 
in general) must take into account the effect of the cosmic expansion on 
photon momentum, so this equation takes the form
\eqa
{\frac{dt}{d{\lambda}}}{\frac{{\partial}f}{{\partial}t}}+
{\frac{dx^{\mu}}{d{\lambda}}}{\frac{{\partial}f}{{\partial}x^{\mu}}}-
{\Big (}{\topstar{\Gamma}}_{t{\nu}}^{{\,}i}{\frac{dt}{d{\lambda}}}+
{\topstar{\Gamma}}_{{\beta}{\nu}}^{{\,}i}{\frac{dx^{\beta}}{d{\lambda}}}
{\Big )}p_{(t)}^{\nu}{\frac{{\partial}f}{{\partial}p_{(t)}^i}}=
C_{\rm coll}[f],
\ena
where ${\lambda}$ is an affine parameter along the photon path.

Given the complications, it is quite certain that exact treatments of neutrino 
decoupling and primordial nucleosynthesis in quasi-metric cosmology via 
Boltzmann equations will be rather cumbersome. For this reason we 
have used approximate methods instead. However, even approximate methods
should be good enough to make rough predictions testable against observations.
\end{document}